\documentclass[prb,aps,nobibnotes,twocolumn]{revtex4}
\usepackage{graphicx}%
\usepackage{dcolumn}
\usepackage{amsmath}   
\usepackage{bm}
\usepackage{dcolumn}
\usepackage{longtable}
\begin{document}

\title{\centering\Large\bf Thermodynamics  and Dynamics of a
 Monoatomic Glass-Former. Constant Pressure and Constant Volume Behavior }
\author{Vitaliy  Kapko} 
\author{Dmitry V.\ Matyushov} 
\affiliation{Center for Biological Physics, Arizona State University, 
PO Box  871504, Tempe, AZ 85287-1504}  
\author{C. Austen\ Angell} 
\affiliation{Department of Chemistry and Biochemistry, 
  Arizona State University, PO Box 871604, Tempe, AZ 85287-1604}
\date{\today}
\begin{abstract}
  We report constant-volume and constant-pressure simulations of the
  thermodynamic and dynamic properties of the low-temperature liquid
  and crystalline phases of the modified Stillinger-Weber (mSW) model.
  We have found an approximately linear increase of the effective
  Gaussian width of the distribution of inherent structures. This
  effect comes from non-Gaussianity of the landscape and is consistent
  with the predictions of the Gaussian excitations model representing
  the thermodynamics of the configurational manifold as an ensemble of
  excitations, each carrying an excitation entropy. The mSW model
  provides us with both the configurational and excess entropies, with
  the difference mostly attributed to vibrational anharmonicity.  We
  therefore can address the distinction between the excess
  thermodynamic quantities often used
  in the
  Adam-Gibbs (AG) equation. 
  We find a new break in the slope of the
  constant pressure AG plot
  when the excess entropy is
  used in the AG equation. 
  The simulation diffusivity data are equally well
  fitted by applying a new equation, derived within the Gaussian
  excitations model, that emphasizes enthalpy over entropy as the
  thermodynamic control variable for transport in viscous liquids.  
\end{abstract}
\maketitle

\section{Introduction}
\label{sec:1}
Both thermodynamic and dynamic properties of structural glass-formers
are unusual and not fully understood.\cite{Ngai:00,Angell:95} It has
long been suggested that puzzles of the dynamics of supercooled
liquids can be unraveled from the properties of their energy
landscapes.\cite{Goldstein:69} Attempts to build a description of
dynamics based on the liquid thermodynamics go back to the Adam-Gibbs
(AG) theory\cite{Adam:65} which emphasizes configurational entropy as
the origin of non-Arrhenius dynamics. Even though the entropic
paradigm seems to be currently most successful in describing
relaxation,\cite{Xia:00,Xia:01,Lubchenko:04} other approaches have
emphasized the energetic aspects of the problem in terms of the
activated kinetics over enthalpic barriers increasing in height with
lowering
temperature.\cite{Goldstein:69,Baessler:87,Arhipov:94,Dyre:95,DMjcp5:05}
Some very recent data\cite{Dalle-Ferrier:07} seem to disfavour the
existence of diverging lengthscale assumed in ``cooperative region''
models\cite{Adam:65,Xia:00} supporting instead the picture of
activation-based dynamics in glass-forming materials.

Configurational entropy is a property not directly accessible from
laboratory experiment, but is increasingly available from computer
simulations of model
systems.\cite{Sastry:01,Voivod:01,Mossa:02,Giovambattista:03,Voivod:04,Gebremichael:05}
Excess entropy, $s^{\text{ex}}_{P,V}(T)$, i.e.\ the liquid entropy over that
of the thermodynamically stable crystal, has been successfully
used\cite{Angell:91,Richert:98} instead of configurational entropy
$s^c_{P,V}(T)$ in the AG relation
\begin{equation}
  \label{eq:2}
  \ln(\tau_{P,V}(T)/ \tau_0) = \frac{\Delta}{Ts^c_{P,V}(T)} , 
\end{equation}
where $\tau_{P,V}(T)$ is the time of structural relaxation and $\tau_0$ is
the time characteristic of quasi-lattice vibrations. The subscripts
$P,V$ in the entropy and relaxation time specify the conditions,
constant-pressure ($P$) or constant-volume ($V$), at which the entropy
and the relaxation time are measured.  Historically, the definition of
the configurational entropy used by Adam and Gibbs\cite{Adam:65} was
that of the full configurational entropy including the entropy of
basin vibrations. However, since they explicitly demanded the
vibrational entropy to cancel between the liquid and the crystal,
their configurational entropy is in fact the entropy of inherent
structure introduced by Stillinger and Weber (see
below).\cite{Stillinger:82} We will therefore refer to the
configurational entropy in this latter definition.

The full configurational entropy and Stillinger's entropy of inherent
structures are in fact not equivalent.  Most studies, both
laboratory\cite{Goldstein:76,Phillips:89,Corezzi:04} and
computational,\cite{Sastry:01,Mossa:02,Giovambattista:03} have shown
that the excess entropy has a significant contribution from the
vibrational manifold related to the excess density of vibrational
states in liquid, $g_{P,V}^{\text{liq}}(\omega)$, compared to the crystal,
$g_{P,V}^{\text{cryst}}(\omega)$. The excess vibrational entropy from
harmonic motions is a sum over the vibrational spectrum:
\begin{equation}
  \label{eq:1}
     \Delta s^{h}_{P,V} = \sum_{\omega} \left[g^{\text{liq}}_{P,V}(\omega) 
                          - g^{\text{cryst}}_{P,V}(\omega) \right]\ln(\omega) .
\end{equation}
So long as the structure does not change, the excess density of states
is independent of temperature for purely harmonic vibrations resulting
in a temperature-independent excess harmonic entropy $\Delta s_{P,V}^h$ and
zero contribution to the excess heat capacity. The anharmonicity of
atomic and molecular vibrations leads to two effects: (i) sample
expansion at constant-pressure heating and (ii) deviations of the
vibrational excess entropy $\Delta s^{\text{vib}}_{P,V}$ from the harmonic
formula in Eq.\ (\ref{eq:1}).  The first effect makes the density of
quasi-harmonic vibrations vary with temperature and the second effect
requires extracting the thermodynamics of vibrations without relying
on the harmonic approximation [Eq.\ (\ref{eq:1})].
 
For most systems studied to date, the density of quasi-harmonic
vibrations is known to depend weakly on temperature except for the
low-frequency feature known as the Boson peak. The excess vibrational
density of states from this region in liquid selenium was shown to
produce ca.\ 30\% of both the excess entropy and heat
capacity.\cite{Phillips:89} These results, and the puzzling ability of
both the configurational (simulations) and excess (laboratory)
entropies to fit the relaxation data according to Eq.\ (\ref{eq:2}),
have lead to the suggestion that configurational and excess entropies
of glass formers might be proportional to each
other.\cite{Martinez:01,Angell:02} There are simulation\cite{Starr:03}
and laboratory\cite{Corezzi:04} data in support of this proposal,
although the question is still not fully settled.\cite{Johari:07}

\begin{figure}
  \centering
  \includegraphics*[width=7cm]{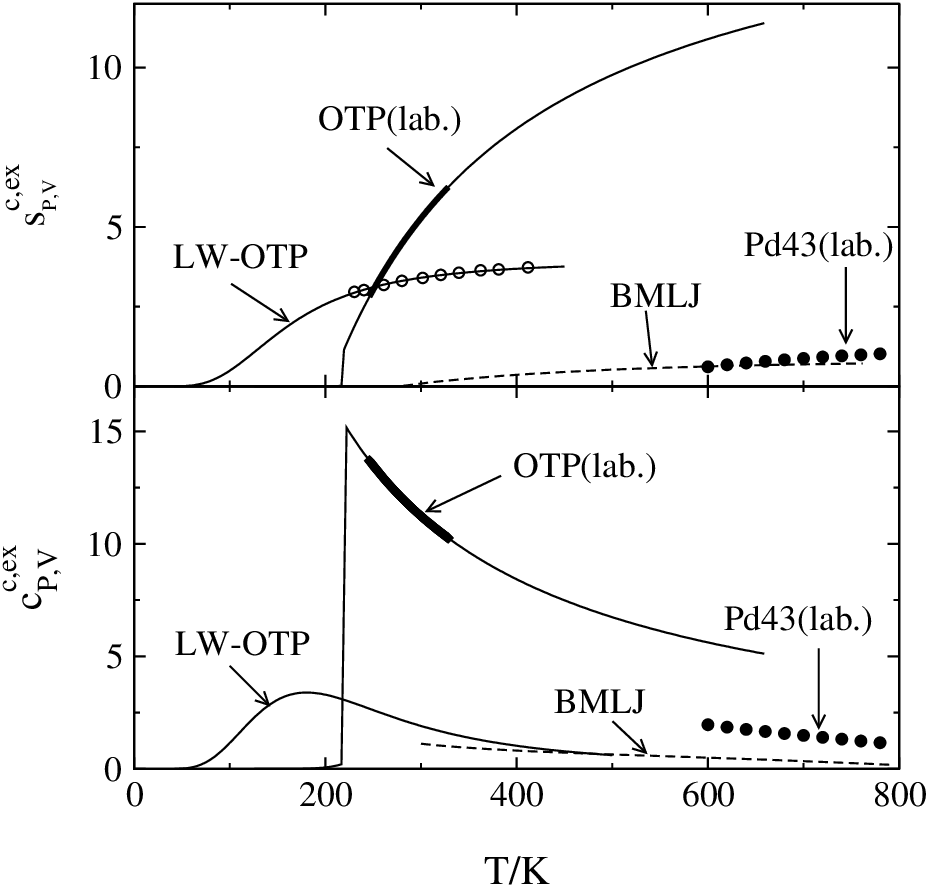}
  \caption{Excess (for laboratory data) and configurational (for
    computer simulations) entropy (a) and heat capacity (b) of some
    glass-forming liquids: glass-forming alloy
    Pd$_{43}$Cu$_{27}$Ni$_{10}$P$_{20}$ (Pd43),\cite{Lu:02}
    o-terphenyl\cite{Moynihan:00} (OTP) and their computer models,
    binary mixture Lennard-Jones (BMLJ)\cite{KobAndersen:94} and Lewis
    and Wahnstr\"om o-terphenyl (LW-OTP).\cite{Mossa:02} The closed
    points and thick lines are laboratory data, the open points and
    dashed lines are computer simulations, and the thin lines are fits
    to the 1G excitations model (from Ref.\ \onlinecite{DMjcp1:07}).
    }
  \label{fig:1}
\end{figure}

The resolution of the problem of partitioning the excess entropy and
heat capacity between vibrational and configurational manifolds is not
straightforward. Recent experimental data have argued in favor of both
a negligible contribution of vibrations to the excess heat capacity
\cite{Johari:07} and a significant contribution amounting from 30\%
(Refs.\ \onlinecite{Phillips:89} and \onlinecite{Corezzi:04}) to
50--60\% (Refs.\ \onlinecite{Angell:03} and \onlinecite{Richert:07}).
Computer models do not tend to help much in resolving this issue due
to a general disconnect between simulated and experimental
results. The situation is illustrated in Fig.\ \ref{fig:1} where
excess (laboratory experiment) and configurational (computer
simulations) thermodynamic quantities are compared for two substances,
$o$-terphenyl and the glass-forming alloy
Pd$_{43}$Cu$_{27}$Ni$_{10}$P$_{20}$, for which both force-field
models\cite{LewisW:94,KobAndersen:94} and experimental
results\cite{Moynihan:00,Lu:02} are available. For $o$-terphenyl,
there is a major difference in both the magnitude and temperature
dependence of the entropy and heat capacity, though this is not
unexpected given that a flexible 14-carbon molecule is being modeled
with a rigid three-bead particle (making it much less entropic).  The
difference is not removed by comparing the experiments with the
limited simulation data available for constant
pressure.\cite{Angell:03} For Pd$_{43}$Cu$_{27}$Ni$_{10}$P$_{20}$
glass-former, as modeled by Kob and co-workers binary
mixture,\cite{KobAndersen:94} the agreement is better and a noticeable
difference exists only for the heat capacities, however the need for
many components in the experimental analog is a disadvantage.

Clearly, to resolve the question of vibrational contributions to the
excess heat capacity, the field is in need of better model systems.  A
minimum need is for a system in which both excess and configurational
data are available, something so far lacking in computer models
(except for water\cite{Starr:03} which is famously anomalous and
unrepresentative of the glass problem we are addressing).  The recent
exploration of the Stillinger-Weber type model by Molinero \textit{et
  al.}\cite{Molinero:06} offers an opportunity to fill this gap since
the crystal state is available and, for chosen parametrizations, the
liquid can be supercooled without limit. Our simulations here take
advantage of this opportunity.

\begin{figure}
  \centering
  \includegraphics*[width=7cm]{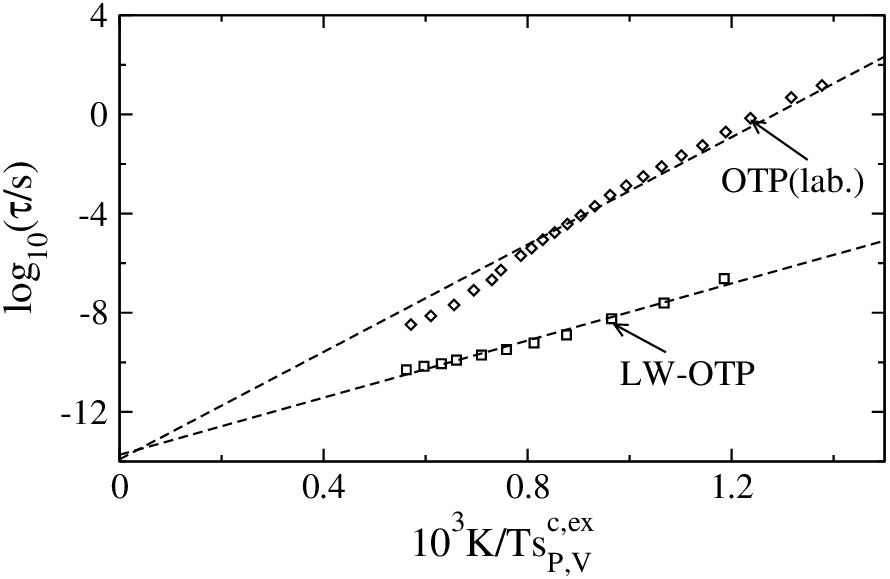}
  \caption{Adam-Gibbs plot of the experimental dielectric relaxation
    data of $o$-terphenyl (OTP)\cite{Richert:98} and simulated
    self-diffusivity\cite{Mossa:02} of Lewis and Wahnstr\"om o-terphenyl
    (LW-OTP, NVT simulations). Excess entropy (at constant pressure)
    was used for experimental points and configurational entropy (at
    constant volume) for simulated data. The dashed line is a linear
    fit through the data with the intercept at $\tau_0=10^{-14}$ s. The
    simulation data are shifted vertically to cross the vertical axis
    at about the same relaxation time. The Adam-Gibbs relation does
    not linearize the dielectric data in the whole range of
    temperatures displayed,\cite{Richert:98} and the plot only serves
    to illustrate the difference in the range of Adam-Gibbs parameter
    $(Ts^{c,\text{ex}}_{P,V}(T))^{-1}$ and relaxation times explored by
    simulations ($V=Const$) and experiment ($P=Const$).}
  \label{fig:2}
\end{figure}

Many recent simulations have offered access to both configurational
thermodynamics and dynamics reporting the success of the AG
relation\cite{Scala:00,Sastry:01,Mossa:02,Gebremichael:05} [Eq.\
(\ref{eq:2})].  As mentioned above, the low-temperature part of
experimental relaxation data can be fitted by the same relation where
$s^{\text{ex}}_{P,V}(T)$ is used instead of
$s^c_{P,V}(T)$.\cite{Richert:98} While this creates a puzzling
contradiction, one needs to keep in mind the difference in the scales
of the two sets of data, which is illustrated in Fig.\ \ref{fig:2}
comparing the laboratory dielectric data\cite{Richert:98} for OTP with
the results of simulations\cite{Mossa:02} for the Lewis and Wahnstr\"om
(LW) model of OTP.\cite{LewisW:94} The difference in slopes of
simulations and laboratory data may originate from the use of
different ensembles, constant volume and constant pressure,
respectively, and of different entropies, configurational and
excess. The AG relation in fact linearizes the laboratory data only at
low temperatures, below the crossover temperature $T_x$ associated
with either the mode-coupling critical temperature or the temperature
$T_b$ of the Stickel analysis.\cite{Richert:98} The experimental data
can be linearized in the AG plot with different slopes below and above
$T_x$, as we also describe below for the modified Stillinger-Weber
(mSW) model, when the excess entropy is used in the AG plot.

\begin{figure}
  \centering
  \includegraphics*[width=7cm]{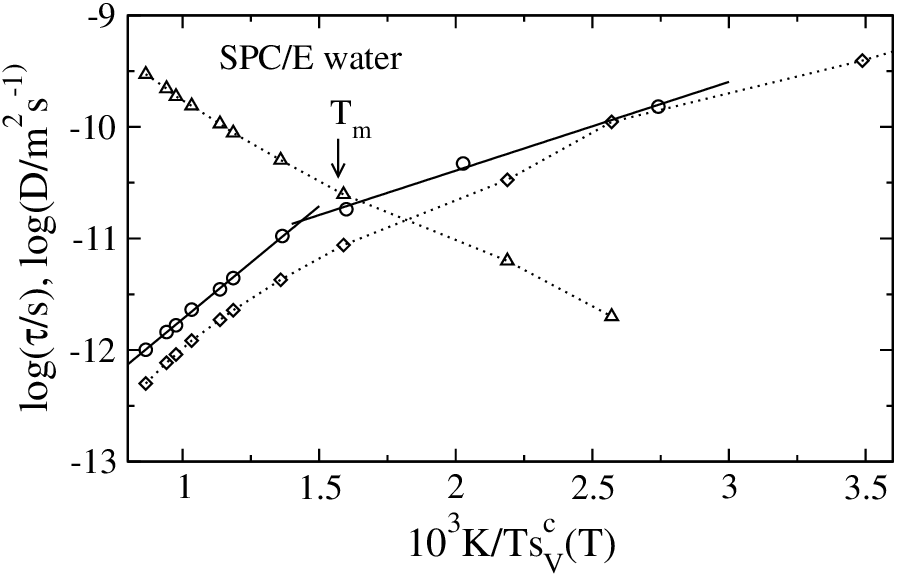}
  \caption{Diffusivity (triangles), relaxation time of one-particle
    rotational diffusion (diamonds), and Debye relaxation time
    (circles) of SPC/E water\cite{DMjpcb1:06} vs $1/(Ts^c_V(T))$ with
    configurational entropies taken from Ref.\ \onlinecite{Starr:03}.
    The dotted lines connect the points, the solid lines are linear
    regressions through high- and low-temperature portions of the
    dielectric relaxation data.  $T_m=273$ K indicates the melting
    temperature of laboratory water at ambient conditions.}
  \label{fig:3} 
\end{figure}

Both the range of temperatures and the property studied can affect the
conclusions regarding the validity of the AG relation from the
simulation data. This is illustrated for SPC/E water in Fig.\
\ref{fig:3}. The dielectric relaxation times collected from Molecular
Dynamics (MD) simulations in a broad range of
temperatures\cite{DMjpcb1:06} show a break in the slope of the AG plot
(configurational entropy is taken from Ref.\
\onlinecite{Starr:03}). The change in slope is much less pronounced
for the one-particle rotational and translational relaxation times
(triangles and diamonds in Fig.\ \ref{fig:3}) than for the
many-particle Debye relaxation time (circles in Fig.\ \ref{fig:3}).
Diffusivity was the only dynamic property considered in the original
report of success of the AG relation for SPC/E water,\cite{Scala:00}
which indeed shows almost linear AG plot.  Notice that the change of
the slope seen in the simulation data in Fig.\ \ref{fig:3} is
uniformly observed in laboratory dielectric measurements of a number
of glass-formers.\cite{Richert:98} However, a downward break in the
slope, in contrast to the upward-curved dependence seen for SPC/E
water in Fig.\ \ref{fig:3}, is typically obtained for molecular
glass-formers, as is also the case with the AG plot of the mSW model
discussed below.  In case of water, this discussion should be
supplemented by the question of the relevance of the AG plot to the
approach to a thermodynamic (near-critical) singularity.\cite{Xu:05}

Two of us have recently suggested to describe the thermodynamics and
dynamics of glass-forming liquids close to $T_g$ in terms of
configurational excitations with a Gaussian distribution of excitation
energies (Gaussian excitations model).\cite{DMjcp5:05,DMjcp1:07}
Entropy, calculated from microcanonical ensemble, was the starting
point for developing the theory.  Each excitation was assumed to carry
the excess entropy $s_0$ which was subdivided into configurational and
vibrational parts, $s_0 = s_0^{\text{vib}} + s_0^c$. These excitation
entropies and the number of configurations of an ideal gas of
excitations provide the excess and configurational components of the
entropy. The excess entropy is connected to the configurational
entropy by the relation
\begin{equation}
  \label{eq:1-1}
  s_{P,V}^{\text{ex}}(T) = s_{P,V}^c[x(T),T] + x(T) s_0^{\text{vib}} .
\end{equation}
The effect of the vibrational excitation entropy $s_0^{\text{vib}}$ is
not limited to the last term in Eq.\ (\ref{eq:1-1}) since the
excited-state population $x(T)$ is determined by the total excitation
entropy $s_0$ and the excitation energy $\epsilon_0$:
\begin{equation}
  \label{eq:1-2}
 x(T) = \left(1 + \exp\right[s_0 - \beta(\epsilon_0 -2x(T)\lambda)\left] \right)^{-1} ,
\end{equation}
where $\beta$ is the inverse temperature. The energies of excitations are
assumed to belong to a Gaussian distribution with the width $2\lambda T$ and
the Gaussian width parameter $\lambda$. This energy lowers the energy of
excitations in Eq.\ (\ref{eq:1-2}) to the extent determined by the
excited state population and $\lambda$, $\epsilon(T)=\epsilon_0-2x(T)\lambda$, thus producing a
self-consistent equation for $x(T)$.

The excess vibrational entropy originating in the excess density of
states of a liquid relative to its crystal provides an extra driving
force, in addition to the larger number of basins, for the system to
reach the top of its energy landscape.\cite{Goldstein:76} This extra
entropic driving force will lead to an increased fragility of the
liquid if the excited state population change occurs in the
temperature range of interest (near $T_g$). In such a case, changes in
the excess density of vibrational states will be observable directly
through neutron scattering study of glasses of different fictive
temperatures, as has been reported for some cases. However, there are
strong theoretical suggestions\cite{DMjcp5:05,DMjcp1:07} that in the
case of very fragile liquids the excited state is highly populated
above $T_g$ (due, in principle, to an earlier phase transition below
$T_g$\cite{DMjcp5:05,DMjcp1:07}) in which case the excess entropy due
to vibrations (which can be quite
significant\cite{Goldstein:76,Angell:02}) will change little with
temperature. There will be a vibrational entropy difference from the
crystal due to different vibrational density of states [Eq.\
(\ref{eq:1})], but there will be no temperature dependence to this
excess. The higher heat capacity of the fragile liquid must then have
some other source. One such source could lie in the $T$-dependence of
the Gaussian width of the excitation profile, which will be addressed
below.

This conclusion is, of course, limited by the assumption that
$s_0^{\text{vib}}$ is temperature-independent.  This assumption in
fact implies the neglect of vibration anharmonicity making eigenvalues
of the landscape basins depend on temperature, also including the
effect of thermal expansion. If this assumption is
violated,\cite{Angell:03} a vibrational contribution appears in the
excess heat capacity. We found, however, that the assumption of
$s_0^{\text{vib}}=Const$ describes laboratory fragile liquids well and
also found that common model fluids used in simulations classify as
intermediate/strong in terms of their
fragility.\cite{DMjcp5:05,DMjcp1:07}

For intermediate and strong liquids, the temperature change in the
configurational entropy is driven by the changing population of the
configurationally excited state, as was the case in the original
two-state Angell-Wong model.\cite{Angell:70} In this case, the
temperature dependence of $x(T)$ in Eq.\ (\ref{eq:1-1}) contributes to
the heat capacity, and there is a non-zero vibrational component in
the excess heat capacity.\cite{Angell:02} A nonzero $s_0^{\text{vib}}$
is therefore sufficient to produce a vibrational excess heat capacity
for intermediate/strong liquids, while this property should be
temperature-dependent for a vibrational excess heat capacity to exist
for fragile liquid.

The distinction between the thermodynamics of fragile and strong
glass-formers is attributed in the Gaussian excitations model to two
parameters: critical excitation entropy and critical temperature.  In
order for the fragile behavior to be realized, the excitation entropy
should be higher than the critical value $s_{0c}\simeq 2$ and the
temperature should be lower than $T_c=\lambda/2$. The binary mixture Lennard-Jones (BMLJ)
liquids, which we have analyzed previously,\cite{DMjcp1:07} and the
mSW liquid analyzed here all have excitation entropies falling below
the critical value and thus classify as strong/intermediate liquids.
The excess heat capacity of the mSW liquid has, therefore, a
vibrational component $\simeq 15$ \% as discussed below.

The dynamic part of the excitation model\cite{DMjcp1:07} has offered
an expression for the relaxation time alternative to the AG formula in
terms of the configurational heat capacity instead of the configurational
entropy:
\begin{equation}
  \label{eq:3}
   \ln(\tau/ \tau_0) = \frac{DT'}{T-T'c_P^c(T)} , 
\end{equation}
where $D$ and $T'$ are model parameters varied in fitting the
experiment.  This relation gives the fits of relaxation times from
laboratory and computer experiments comparable to those based on the
AG relation.  However, it carries a problem similar to the one
encountered in applications of the AG equation.  The experimental data
can be fitted by using the excess heat capacity, $c_P^{\text{ex}}$,
while simulation data can be accounted for with the configurational
part $c_P^{c}$. Here we offer some insights into this problem from the
data collected for the mSW model.\cite{Molinero:06}

Some predictions of the excitations model deviate from popular models
of landscape thermodynamics. The concept of potential energy
landscape, proposed by Goldstein,\cite{Goldstein:69} was formalized by
Stillinger and Weber\cite{Stillinger:82} in terms of the
thermodynamics of inherent structures reducing the many-body problem
of describing liquids to a single ``reaction coordinate'' $\phi$ defined
as the depth of the potential energy minimum (per molecule, $\phi=\Phi/N$,
$N$ is the number of molecules). The probability to find a minimum of
depth $\phi$ is determined by the number of minima of given depth $\Omega(\phi)$
and two free energies, the total thermodynamic free energy $f(T)$ and
the free energy $f^b(\phi,T)$ of the system exploring the phase space
within the basin surrounding the minimum of depth $\phi$:
\begin{equation}
  \label{eq:4}
  N^{-1} \ln\left[ P(\phi,T)\right] = -\beta \phi + \omega(\phi) + \beta\left[f(T) - f^b(\phi,T)\right] .
\end{equation}
Here, $\omega(\phi) = N^{-1}\ln[\Omega(\phi)]$ is
the enumeration function. The subscript specifying ensemble (constant
volume or constant pressure) is omitted in Eq.\ (\ref{eq:4}) for
brevity. In case of constant-volume conditions, $f(T)$ is the common
notation for the free (Helmholtz) energy per particle. At constant
pressure, $f(T)$ should be understood as the Gibbs energy and $\phi$ is
the potential enthalpy minimum.\cite{StillingerJPCB:98} We will drop
the ensemble specification in the landscape variables below
reserving it to the equilibrium properties, e.g. $c_{P,V}^c(T)$, where
this distinction is critical.

The formalism of inherent structures is particularly convenient when
$f^b(\phi,T)$ is independent of $\phi$.  Otherwise, Eq.\ (\ref{eq:4}) is
formally a definition of the basin free energy. It was found that at
high temperatures accessible to simulations the harmonic part of
$f^b(\phi,T)$ is a weak linear function of $\phi$.\cite{Starr:01,Mossa:02}
The main focus of the formalism is, however, on the enumeration
function. It is a decreasing function with lowering $\phi$ allowing the
liquid to explore higher energy states with increasing temperature
(entropy drive). It was suggested that Derrida's Gaussian
model,\cite{Derrida:81} originally derived for glasses with quenched
disorder,\cite{Fischer:99} can be extended to quasi-equilibrated
supercooled liquids with the result that $\omega(\phi)$ is an inverted
parabola with a temperature-independent curvature:
\begin{equation}
  \label{eq:5}
  \omega(\phi) = \omega_0 - \frac{(\phi-\phi_0)^2}{2\sigma^2} .
\end{equation}

Even though combinatorial arguments suggest that the parabolic
approximation should fail at low
temperatures,\cite{Stillinger:88,Shell:04} the high-temperature
portion of $\omega(\phi)$ is supported by simulations of Lennard-Jones (LJ)
mixtures.\cite{Buchner:99,Sastry:01} A more stringent test of the
Gaussian model comes from considering the temperature dependence of
the average basin depth $\bar \phi_V(T)$ which, in the Gaussian model, is
predicted to scale linearly with $1/T$, producing a $1/T^2$ scaling
for the configurational heat capacity.  Most available simulations
report deviations from this
scaling.\cite{SastryJP:00,Sciortino:05,Moreno:06,DMpre:07} It is
currently not fully established whether the origin of this deviations
should be traced back solely to anharmonicity
effects\cite{SastryJP:00} or to the actual failure of the Gaussian
approximation although two recent simulations exploring the
low-temperature part of the landscape point to the latter
explanation.\cite{Moreno:06,DMpre:07} In the excitations model, the
temperature dependence of the average basin energy is qualitatively
different for fragile and intermediate liquids. In the former case,
the basin energy is essentially flat for fragile liquids terminating
through a discontinuity at the phase transition below $T_g$. For
intermediate liquids, $\bar\phi_V(T)$ starts to dip as $1/T$ from a
high-temperature plateau (as was found for the BMLJ
liquids\cite{Sastry:01}) and then inflects into an exponential
temperature dependence recently observed in simulations of model
network glass-formers.\cite{Moreno:06} Note that the excitations model
is the only analytical model we are aware of which incorporates both
types of temperature scaling in one formalism.

The Gaussian excitations model leads to a non-Gaussian $P(\phi)$ and thus
a non-parabolic $\omega(\phi)$.\cite{DMjcp5:05} When the non-Gaussian
distribution $P(\phi)$ is fitted to a Gaussian function, the result is an
approximately linear scaling of the squared width with temperature
$\sigma_{P,V}^2(T) \propto T$.  This results makes $P(\phi)$ in Eq.\ (\ref{eq:4}) a
Boltzmann distribution, which seems to be more relevant for a
(quasi)equilibrated supercooled liquid than the non-Boltzmann
distribution of the Gaussian landscape more relevant for systems with
quenched disorder. In addition, the excitations model gives hyperbolic
temperature scaling for the heat capacity $c_{P,V}^c \propto 1/T$. This
scaling is often observed in the laboratory for $c_P^{\text{ex}}(T)$,
but here we again face the same problem as above concerning the
connection between excess and configurational heat capacity.

An approximately linear scaling of the width of $P(\phi)$ with
temperature in the excitations model is the result of the assumed
mean-field, infinite-range attraction between the excitations, which
is equivalent to assuming a Gaussian manifold of real-space excitation
energies with the variance $2\lambda T$. A finite range of interactions
between the excitations will produce a more complex temperature
dependence. For instance, a recent exactly solvable landscape model of
the fluid of dipolar hard spheres\cite{DMpre:07} gave a fairly complex
temperature scaling of the distribution variance $\sigma_{V}^2(T) \propto (1 + \beta
b)^{-3}$ ($b$ is an interaction parameter).  Notice in this regard
that a fluid with dipolar interactions is an archetypal system in
which the mean-field approximation is not applicable.  The reason is
two-fold: (i) the average of the potential is zero and fluctuations is
the first non-vanishing contribution to the
thermodynamics\cite{Osipov:97} (cf.\ to LJ forces described reasonably
by the mean-field van der Waals model) and (ii) the interaction
potential is strongly anisotropic. How the model should be extended to
a finite range of correlations between the excitations is not clear
now, but the distribution width is expected to transform to a
temperature-independent value for isotropic short-range LJ forces, in
compliance with the Gaussian model [Eq.\ (\ref{eq:5})]. Any
non-Gaussian landscape will generate a temperature-dependent width
when distribution of inherent energies is fitted by a Gaussian.

\begin{table*}
  \centering
  \caption{Thermodynamic parameters (energies in kelvin and entropies in $k_{\text{B}}$ units) 
           of mSW and  BMLJ models.}
\label{tab:1}
\begin{ruledtabular}  
\begin{tabular}{lcccccccccc}
Model          & $\epsilon$\footnotemark[1] & $T_m$  & $T_K^c$\footnotemark[2] & $T_K^{\text{ex}}$\footnotemark[3]  & $T_0$\footnotemark[4] &$-\phi_{\text{IG}}$\footnotemark[5] & 
$-\phi_{\text{DC}}$\footnotemark[6] & $\omega_0$\footnotemark[7] & $s_0^{\text{ex}}$ \footnotemark[8] &
$s_0^c$ \footnotemark[9] \\
\hline
mSW ($\lambda=19$)   & 25150  & 806  & 434   &  392 & 420 & 49482 & 50300  &  1.20 & 2.4 & 1.0\\
BMLJ ($\rho=1.2$) & 120    &      & 35    &      &   &   &       &  0.93 &   & 0.32   \\
\end{tabular}
\footnotetext[1]{Lennard-Jones energy of the mSW and BMLJ potentials.}
\footnotetext[2]{From extrapolating $s^c_P(T)$ to zero for the mSW liquid
  and from the constant-volume data\cite{Sastry:01} for the BMLJ
  liquid. $T_K^c=487$ K obtained from $s^c_V(T)$. }
\footnotetext[3]{From solving the equation $s^{\text{ex}}_{P}(T)=0$ with excess entropy
given by Eq.\ (\ref{eq:22}).}
\footnotetext[4]{From the fit of diffusivity data from MD simulations to
the Vogel-Fulcher-Tammann equation.}
\footnotetext[5]{Extrapolated from $\bar\phi_P(T)$ obtained from
NPT simulations to the temperatures $T_K^c$ at which $s_P^c(T)$
becomes zero; $-\phi_{IG}=49265$ K for the NVT ensemble.}
\footnotetext[6]{Basin depth of diamond cubic crystal.}
\footnotetext[7]{Top of the enumeration function corresponding to $T\to \infty$ limit 
 [e.g., $\omega_0$ in Eq.\ (\ref{eq:5})].
  For mSW fluid the number was obtained from fitting the simulation
  data for $s_c(T)$ by a polynomial in $1/T$; for BMLJ fluid the value
  listed is from Ref.\ \onlinecite{Sastry:01}. } 
\footnotetext[8]{Obtained by numerical extrapolation the simulated excess entropy $s^{\text{ex}}_P(T)$ of the
  mSW fluids over the DC crystal to the limit $T\to\infty$. }
\footnotetext[9]{Configurational component of the excitation entropy obtained from
fitting the configurational thermodynamics data from numerical simulations.}
\end{ruledtabular}
\end{table*}

In this paper, we use the results of simulations of the mSW potential
to address some of the challenges listed above. We present the results
for the landscape thermodynamics for two values of the tetrahedrality
parameter $\lambda$ of the mSW potential (see below) and will compare the
results of the analysis to some other models of glass-formers on one
hand and to laboratory data for metallic glass-formers on the
other. Some thermodynamic parameters relevant to our discussion are
listed in Table \ref{tab:1}. All energies throughout below are in
kelvin and entropies and heat capacities are in units of
$k_{\text{B}}$. Also, we use low-case letters for thermodynamic
potentials and energies per liquid particle, e.\ g.\ $s^c_{P,V}(T)$
refers to the configurational entropy per particle.

\section{Landscape Thermodynamics}
\label{sec:2}
The properties of the energy landscape for a given interaction
potential can be studied by either a direct calculation of the
enumeration function $\omega(\phi)$ or by looking at the ensemble
averages.\cite{StillingerJPCB:98} In the first route, $\omega(\phi)$ is
calculated by patching together the distribution functions at
different temperatures once the total and basin free energies entering
Eq.\ (\ref{eq:4}) are known.\cite{SastryJP:00,Sciortino:05,DMpre:07}
The width of each individual distribution scales as $1/ \sqrt{N}$ with
the number of particles $N$, and most simulation data allow a Gaussian
fit
\begin{equation}
  \label{eq:6}
  P(\phi) \propto e^{-N\left(\phi-\bar \phi_{P,V}(T)\right)^2/2\Gamma_{P,V}(T)^2} .
\end{equation}
Here, $\Gamma_{P,V}(T)$ is an empirical Gaussian width and the
stationary point $\bar\phi_{P,V}(T)$ is the solution of the equation
\begin{equation}
  \label{eq:7}
  \beta\left(1 + \partial f^b(\phi,T)/ \partial \phi \right) = \partial \omega(\phi)/ \partial \phi .
\end{equation}
When $f^b(\phi,T)=f^b(T)$ is independent of $\phi$ (harmonic approximation)
and the enumeration function is given by the inverted parabola [Eq.\
(\ref{eq:5})], one gets the hyperbolic temperature scaling
\begin{equation}
  \label{eq:8}
  \bar\phi_{P,V}(T) = \phi_0 - \frac{\sigma^2}{T} 
\end{equation}
characteristic of the Gaussian landscape. 

The configurational entropy is determined by the enumeration function
taken at the average basin depth
\begin{equation}
  \label{eq:8-1}
  s^c_{P,V}(T) = \omega(\bar\phi_{P,V}(T)). 
\end{equation}
The configurational entropy is then a part of the thermodynamic free energy
$f(T)$
\begin{equation}
  \label{eq:8-2}
  f(T) = \bar \phi_{P,V}(T) - Ts^c_{P,V}(T) + f^b[\bar\phi_{P,V}(T),T].  
\end{equation}
Once both $\bar\phi_{P,V}(T)$ and $s^c_{P,V}(T)$ are known, the
enumeration function $\omega(\phi)$ can be calculated using Eq.\
(\ref{eq:8-1}).\cite{StillingerJPCB:98}

\section{Simulation details}
\label{sec:3}
We use a model of network liquids which was originally
introduced by Stillinger and Weber (SW)\cite{StillingerW:85} for  
silicon. In the SW model, a three-body term is added to the 
pairwise potential $v_2(r)$ to introduce penalty for deviating
from tetrahedrality
\begin{equation}
  \label{eq:11}
  v(r_{12},r_{23},r_{31}) = v_2(r_{12}) + \lambda v_3(r_{12},r_{23},r_{31})
\end{equation}
where
\begin{equation}
v_2(r) = \left\{\begin{array}{lll}
 &&   A(B/r^4-1)\exp(r-a)^{-1}), r<a\\
 &&  0, r\geq a
 \end{array} \right.
\end{equation}
with $A=7.049555277$, $B=0.602245584$, and $a=1.8$.  The three body
potential has form
\begin{equation}
\label{eq:12}
v_3(r_{12},r_{13},\theta_{213})=\exp[\gamma(r_{12}-a)^{-1}+\gamma(r_{13}-a)^{-1}]\times(3\cos \theta_{213}+1)^2/9
\end{equation}
with $\gamma=1.2$.  The potentials are given in reduced units $\sigma=0.20951$
nm and $\epsilon=50$ kcal/mol (see also Table \ref{tab:1}).

The original SW model with $\lambda=21$ describes silicon, but was modified
recently by Molinero \textit{et al.}\cite{Molinero:06} by decreasing
the tetrahedrality parameter $\lambda$ (in contrast to its increase
attempted earlier by Middleton and Wales\cite{Middleton:01}) to obtain
monoatomic glass-formers. They showed that the system crystallizes
into diamond cubic (DC) lattice for $\lambda>20.25$ and into body-centered
cubic (BCC) lattice for $\lambda<17.5$.  For intermediate values of $\lambda$ the
fluid fails to crystallize on the time-scale of computer simulation,
producing glass-formers.

Since each fluid, characterized by a given value of $\lambda$, has an
equilibrium crystalline phase, this property can be used to obtain
both the excess and configurational thermodynamics for the same
system.  Two fluids have been used in simulations: the original SW
model ($\lambda=21$) and mSW model ($\lambda=19$).  The results were obtained from
NVT and NPT MD simulations using the constraint method\cite{Allen:96}
and its modification for the NPT ensemble.\cite{BrownClarke:84}
Periodic boundary conditions for the cubic cell of 512 particles were
applied and the time step was about 1.53 fs.  The temperature has been
changed in step-like way with 50 K (0.002$\epsilon$) per jump.  The run
length at each given temperature varies between 0.76 ns at high
temperatures to up to 3 ns at the lowest temperature equivalent to
cooling rates of 65 K/ns and 16 K/ns, respectively.

\begin{figure}
  \centering
  \includegraphics*[width=7cm]{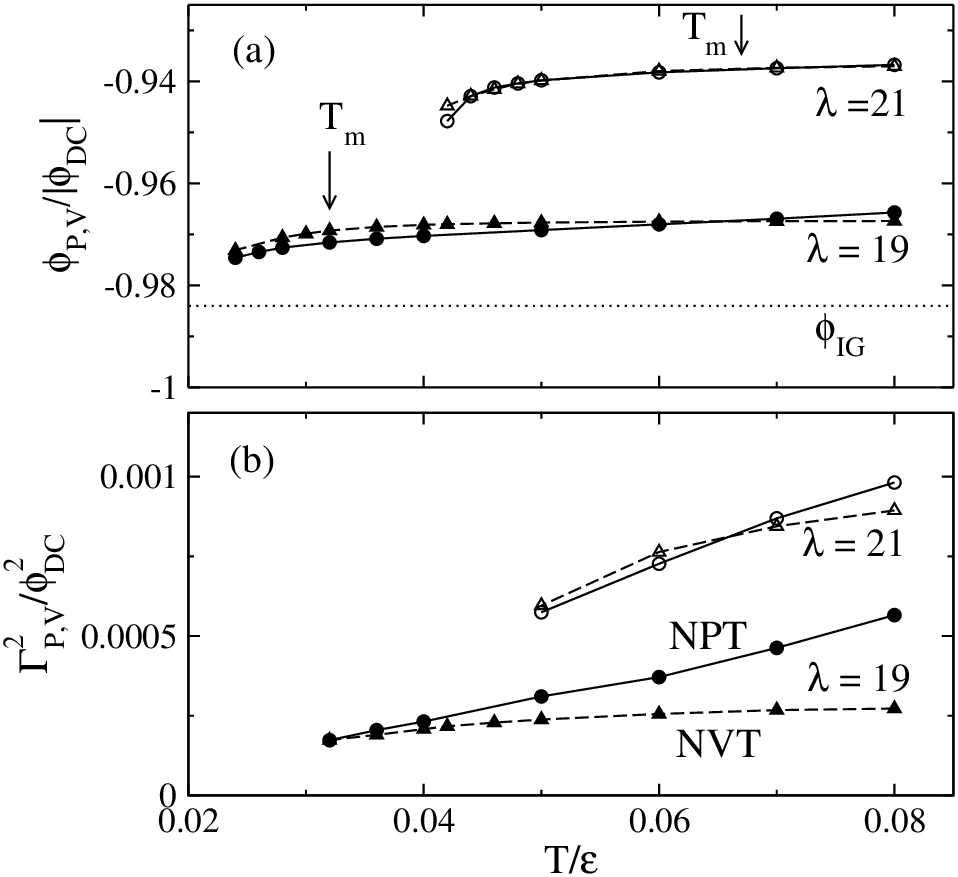}
  \caption{Average basin energy (a) and the effective Gaussian width
    [(b), Eq.\ (\ref{eq:6})] vs temperature reduced by the LJ energy
    $\epsilon$ (Table \ref{tab:1}).  Open points refer to $\lambda=21$, closed
    points indicate $\lambda=19$.  Triangles and circles refer to NVT and
    NPT simulations, respectively. The depth of the diamond cubic
    basin $\phi_{\text{DC}}$ (Table \ref{tab:1}) is used for the energy
    scale. $T_m$ in (a) marks the melting temperature and
    $\phi_{\text{IG}}=\phi_P(T_K^c)$ marks the depth of the ideal-glass
    minimum at which the configurational entropy $s^c_{P}(T)$ becomes
    zero ($\lambda=19$). }
  \label{fig:4}
\end{figure}

\section{Results}
\label{sec:4}

\subsection{Vibrational thermodynamics}
\label{sec:4-1}
The excitation profiles $\bar \phi_{P,V}(T)$ and the distribution widths
$\Gamma_{P,V}(T)$ for two SW potentials characterized by $\lambda=19$ and $\lambda=21$
are shown in Fig.\ \ref{fig:4}.  The average basin depth $\bar\phi_{P,V}(T)$
from simulations was fitted to the function
\begin{equation}
  \label{eq:22-2}
  \bar\phi_{P,V}(T) = a_{P,V}^{(0)} + a_{P,V}^{(1)} T^* + 
                   a_{P,V}^{(2)} (T^*)^2 + b_{P,V}^{(1)}/T^* + b_{P,V}^{(2)}/(T^*)^2 
\end{equation}
with $T^*=T/ \epsilon$. The expansion coefficients in Eq.\ (\ref{eq:22-2})
are listed in Table \ref{tab:2}.  The basin depth, shown in Fig.\
\ref{fig:4} relative to the equilibrium crystalline (DC) state,
changes little on this energy scale. The most interesting observation
is an approximately linear increase of the effective width with
temperature.  This result is inconsistent with the Gaussian landscape
model [Eq.\ (\ref{eq:5})].  A part of the of the width increase at
constant pressure comes from thermal expansion (cf.\ triangles with
circles in Fig.\ \ref{fig:4}), but there is still an increase of the
effective width by a factor of nearly 2 even in constant-volume
simulations.  Also shown in Fig.\ \ref{fig:4} is the energy of the
deepest amorphous minimum $\phi_P(T_K)$, corresponding to the ideal-glass
transition, measured at the Kauzmann temperature $T_K$ at which the
configurational entropy $s_P^c(T)$ becomes zero (Table \ref{tab:1}).
This minimum lies about 820 K above the crystalline minimum, which can
be compared to Stillinger's estimate of 460 K for laboratory
OTP.\cite{StillingerJPCB:98} The excess values are consistent with the
requirement that even ideal glasses are metastable with respect to the
corresponding crystals.

\begin{table*}
  \caption{Fitting coefficients in Eq.\ (\ref{eq:22-2}) for the average energy of inherent 
       structures obtained from NPT and NVT MD simulations. }   
\label{tab:2}
\centering
 \begin{ruledtabular}  
  \begin{tabular}{lcccccc} 
$\lambda$ & Ensemble & $a_{P,V}^{(0)} $ & $a_{P,V}^{(1)}$ & $a_{P,V}^{(2)}$ &  $b_{P,V}^{(1)}$ & $b_{P,V}^{(2)}$ \\
 \hline
$\lambda=19$ & NPT & $-2.0013$  &  $0.79174$ &  $ -2.2294$ & $0.0020186$ & $-2.86\times10^{-5}$ \\
         NVT & $-1.9744$  &  $0.41355$ &  $-1.5993$  & $0.001723$  &  $-3.03\times10^{-5}$ \\
$\lambda=21$ & NPT  & $-8.0414$ & 66.40      &  $-264.45$  & 0.25165    &  $-3.82\times10^{-3}$ \\
         NVT  & $-2.8242$ & 9.5033     &  $-35.642$  & 0.04262    &  $-7.34\times10^{-4}$ \\
 \end{tabular}
 \end{ruledtabular}  
\end{table*}

\begin{figure}
  \centering
   \includegraphics*[width=7cm]{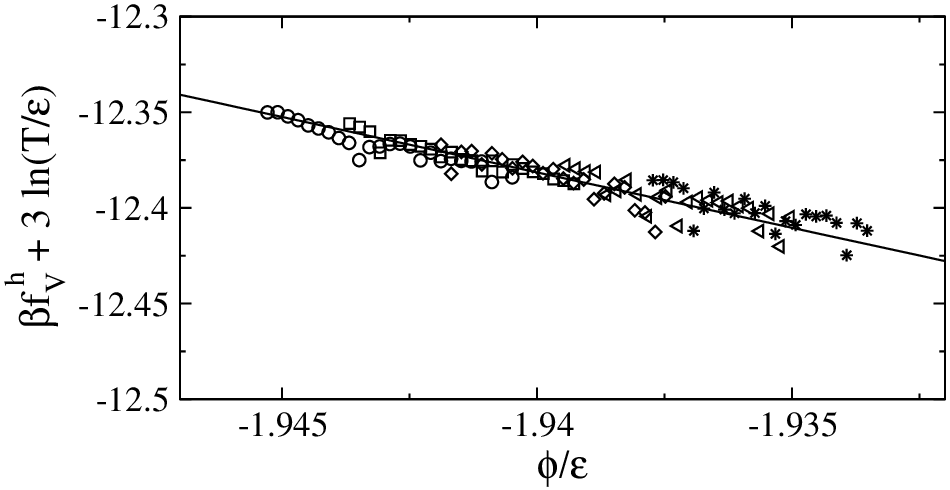}  
   \caption{Basin free energy for the mSW fluid calculated in the
     harmonic approximation [Eq.\ (\ref{eq:13})] from NPT simulations
     at $\lambda=19$, $\rho\sigma^3=0.51$, and varying temperature $T^*=T/ \epsilon$:
     $0.026$ (circles), $0.028$ (squares), $0.03$ (diamonds), $0.036$
     (left triangles), $0.046$ (stars). The solid line is a linear
     regression through the points: $\bar \phi_{P,V}=a_{P,V}+b_{P,V}(\phi/
     \epsilon)$ with $a_P=-23.63$ and $b_P=-5.80$. The NVT data, not shown
     here, are very close to the NPT data with the linear regression
     coefficients $a_V=-22.11$ and $b_V=-5.02$. }
  \label{fig:5}
\end{figure}

In order to gain insight into the origin of the temperature increase
of $\Gamma_{P,V}(T)$ seen in Fig.\ \ref{fig:4} one needs to separate the
basin free energy $f^b(\phi,T)$ in Eq.\ (\ref{eq:4}) from the enumeration
function.  One expects that harmonic approximation holds at low
temperatures when the basin free energy can be obtained by
diagonalizing the Hessian matrix at the local minimum of depth $\phi$
along the simulation trajectory
\begin{equation}
  \label{eq:13}
   \beta f_{P,V}^{h}(\phi,T) = N^{-1} \langle\sum\limits_{i=1}^{3N-3} \ln \beta\hbar\omega_i^{P,V} \rangle_{\phi} .
\end{equation}
We found, as in previous
simulations,\cite{SastryJP:00,Starr:01,Sciortino:05} that $f^{h}_{P,V}(\phi,T)$
obtained from Eq.\ (\ref{eq:13}) is an approximately linear function
of $\phi$ (Fig.\ \ref{fig:5})
\begin{equation}
  \label{eq:14}
 \beta f_{P,V}^{h}(\phi,T) + 3\ln (T/\epsilon) \approx  a_{P,V} + b_{P,V} (\phi/ \epsilon),   
\end{equation}
where the linear regression coefficients are listed in the caption to
Fig.\ \ref{fig:5}. The basins thus become increasingly sharp on
cooling, both at constant pressure and constant volume conditions. The
latter observation is an agreement with the previous constant-volume
simulations of SPC/E water,\cite{Starr:01} but in contrast to the
opposite trend found in constant-volume simulations of BMLJ
fluids.\cite{Sastry:01}

\begin{figure}
   \centering
   \includegraphics*[width=7cm]{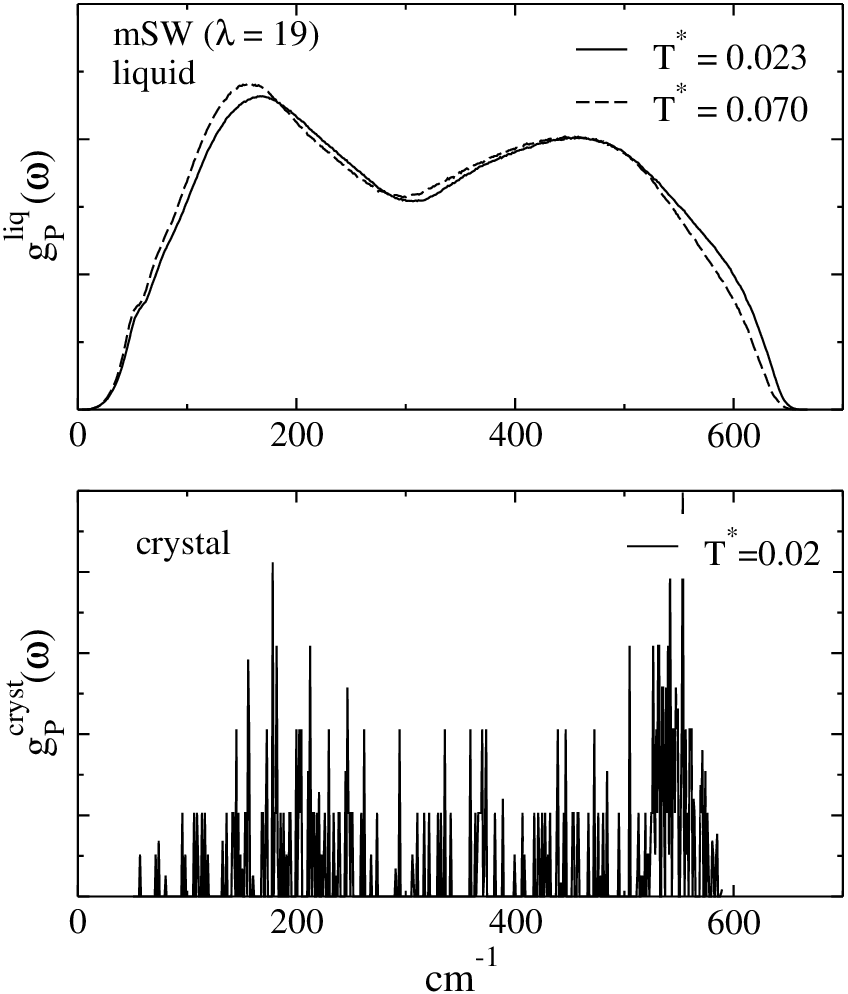}
   \caption{Density of states for mSW model ($\lambda =19$) in the liquid
     state at constant pressure (a) and DC crystal (b); $T^*=T/ \epsilon$,
     where $\epsilon$ is the LJ energy (Tab.\ \ref{tab:1}.) }
   \label{fig:6}
\end{figure}

The vibrational density of states (VDOS) at constant pressure, used to
calculate the harmonic part of the basin free energy, is presented on
Fig.\ \ref{fig:6}. It was obtained by diagonalizing the Hessian of
potential energy in inherent structures using LAPAC's routine
DSYEV.\cite{LAPACK} The VDOS of the liquid phase (Fig.\ \ref{fig:6}a)
is continuous, with two well-defined maxima corresponding to
low-frequency longitudinal and the high-frequency transverse
vibrations.  The VDOS shifts to higher frequencies on cooling, in
agreement with Fig.\ \ref{fig:5}.  The crystalline VDOS was calculated
for the ideal diamond cubic crystal with $N=1728$ particles. In
contrast to the VDOS of the liquid phase, it has a discrete spectrum
sensitive to the system size. Because of this complication, the
vibrational thermodynamics of the crystal was calculated at different
crystal sizes and infinite-size results were obtained by linear
extrapolation of the $1/N$ dependence to $N\to \infty$.

The eigenfrequencies of the Hessian matrix $\omega_i$ depend weakly on
temperature. This effect is caused by anharmonicity of basin
vibrations which tends to soften vibrational frequencies with
increasing temperature. Therefore, in order to properly calculate the
harmonic free energy $f^h_{P,V}(T)$, we used the extrapolation of
$\omega_i(T)$ to $T=0$. In this case, Eq.\ (\ref{eq:13}) with
temperature-independent frequencies gives the expected value for the
harmonic part of the basin internal energy
\begin{equation}
  \label{eq:15}
  e^{h}_{P,V}(T) = \partial(\beta f^{h}_{P,V})/\partial\beta = 3T(1 - 1/N)  .
\end{equation}

Given that the basin free energy $f^h(\phi,T)$ is a linear function of
the basin energy $\phi$ (Fig.\ \ref{fig:5}), the width of the Gaussian
distribution $\Gamma(T)$, obtained by fitting to the probability $P(\phi)$
function to Eq.\ (\ref{eq:6}), is equal to the width $\sigma$ obtained by
quadratic expansion of the enumeration function around the average
basin energy $\bar\phi_{P,V}(T)$ (second derivative of $f(\phi,T)$ in $\phi$ is
zero).  This conclusion is limited by the neglect of the second
derivative of the anharmonic part of the basin free energy
$f^{\text{anh}}(\phi,T)$ which we could not extract from our simulations.
Since the Gaussian landscape precludes temperature dependence of the
width, the approximately linear increase of the width of $P(\phi)$ with
temperature seen in Fig.\ \ref{fig:4} can be assigned to an increase
of the effective width of a non-Gaussian enumeration function fitted
to a Gaussian.\cite{DMpre:07}

\begin{figure}
  \centering
   \includegraphics*[width=7cm]{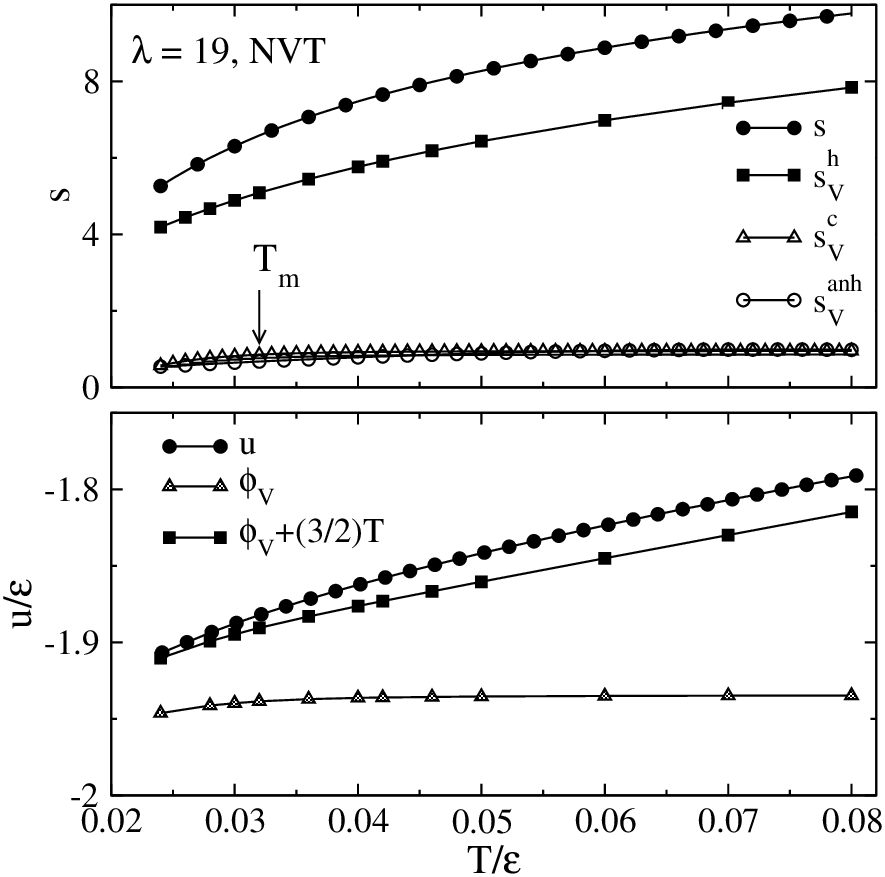}
   \caption{Total entropy $s$ (upper panel) and potential energy $u$
     (lower panel) vs $T$.  The entropy is split into the
     configurational entropy $s^c_V$, harmonic entropy $s_V^{h}$, and
     the anharmonic entropy $s^{\text{anh}}_V$.  The potential energy
     (lower panel, circles) is compared to the average basin depth
     $\bar \phi_V$ (triangles) and the potential energy in the harmonic
     approximation $\bar \phi_V+(3/2)T$ (squares) ($\lambda=19$, NVT
     simulation). }
  \label{fig:7}
\end{figure}

The basins of the mSW fluid are anharmonic, as is seen from the
comparison of the potential energy $u(T)$ to the energy $\bar
\phi_{P,V}(T)+(3/2)T$ (Fig.\ \ref{fig:7}, lower panel). The anharmonic
potential-energy part,
\begin{equation}
\label{eq:15-1}
u^{\text{anh}}_{P,V}(T)=e(T)-3T-\bar{\phi}_{P,V}(T), 
\end{equation}
of the internal energy per particle $e(T)$ can be used to calculate
the anharmonic part of the basin free energy according to the
thermodynamic equation
\begin{equation}
\label{eq:16}
    \beta f^{\text{anh}}_{P,V}(T) = \int\limits_0^T u^{\text{anh}}_{P,V}(T') d\left(\frac{1}{T'}\right) .
\end{equation}

\subsection{Configurational thermodynamics}
Once the harmonic and anharmonic contributions to the basin free
energy are available from Eqs.\ (\ref{eq:13}) and (\ref{eq:16}), the
configurational entropy can be calculated by subtracting the
vibrational (harmonic and anharmonic) entropy of basins from the total
entropy
\begin{equation}
  \label{eq:17}
  s^c_{P,V}(T) = s(T) - s^{h}_{P,V}(T) - s^{\text{anh}}_{P,V}(T) .
\end{equation}
In Eq.\ (\ref{eq:16}), the harmonic entropy is calculated from $T=0$
extrapolated basin frequencies as
\begin{equation}
  \label{eq:18}
  s^{h}_{P,V}(T) = (N)^{-1}\sum_{i=1}^{3N-3}\left[ 1 - \ln(\beta \hbar \omega_i) \right] ,
\end{equation}
and the anharmonic entropy is obtained from Eq.\ (\ref{eq:16}).

Thermodynamic integration\cite{SastryJP:00,Mossa:02} was employed to
calculate the total entropy in Eq.\ (\ref{eq:17}).  The excess free
energy over that of the ideal gas below some reference temperature
$T_r$ was obtained by integrating the internal energy from
simulations:
\begin{equation}
  \label{eq:19}
  \Delta f(T,\rho) = \Delta f(T_r,\rho) + \int_{\beta_r}^{\beta} e(\rho,\beta') d\beta' ,
\end{equation}
where $T_r$ was chosen above the critical temperature. The value $ \Delta
f(T_r,\rho)$ was then obtained by isothermal expansion to the ideal gas
using the equation
\begin{equation}
  \label{eq:20}
  \beta_r \Delta f(T_r,\rho) = \int_0^{\rho} \frac{d \rho'}{\rho'}\left(\frac{ \beta_r P(\rho')}{ \rho'} -1 \right) .
\end{equation}
Finally, the free energy of the ideal gas was added to obtain the
total free energy of the mSW fluid.  

The excess entropy was calculated from the temperature-dependent
enthalpies of the liquid and the crystal according to the relation
\begin{equation}
  \label{eq:22}
  s^{\text{ex}}_P(T) = \Delta s^{\text{fus}}_P + \int_{T_m}^{T} \frac{dT'}{T'} \frac{\partial }{\partial T'}
                  \left(H_{\text{liq}}(T') - H_{\text{DC}}(T')\bigg|_{P} \right) ,
\end{equation}
where $\Delta s^{\text{fus}}_P=2.10$ ($\lambda=19$) is the fusion entropy. The
enthalpies of the liquid phase, $H_{\text{liq}}(T)$, and the DC
crystal, $H_{\text{DC}}(T)$, were fitted from the simulation data to
the following functions:
\begin{equation}
  \label{eq:22-1}
  \begin{split}
    H_{\text{liq}}(T) & = \bar\phi(T) + 3T + c_2 T^2 + c_3 T^3,\\
    H_{\text{DC}}(T) & = -2\epsilon + 3T + d_2 T^2 + d_3 T^3, 
  \end{split}
\end{equation}
where the polynomial coefficients are: $c_2=12.3195 \epsilon^{-1}$, $c_3=-107.993
\epsilon^{-2}$, $d_2=0.9352 \epsilon^{-1} $, and $d_3=26.6414\epsilon^{-2}$.  The configurational
entropy, alternatively to Eq.\ (\ref{eq:17}), can be calculated from
the configurational heat capacity 
\begin{equation}
\label{eq:21}
  s^c_{P,V}(T) = s^c_{P,V}(T_0) + 
           \int\limits_{T_0}^T \frac{1}{T'} \frac{d \bar \phi_{P,V}(T')}{d T'} dT' ,
\end{equation}
where $T_0$ is some temperature for which $s^c_{P,V}(T_0)$ is known from the
thermodynamic integration to the ideal gas.  The two thermodynamic
routes give identical results.

\begin{figure}
  \centering
  \includegraphics*[width=7cm]{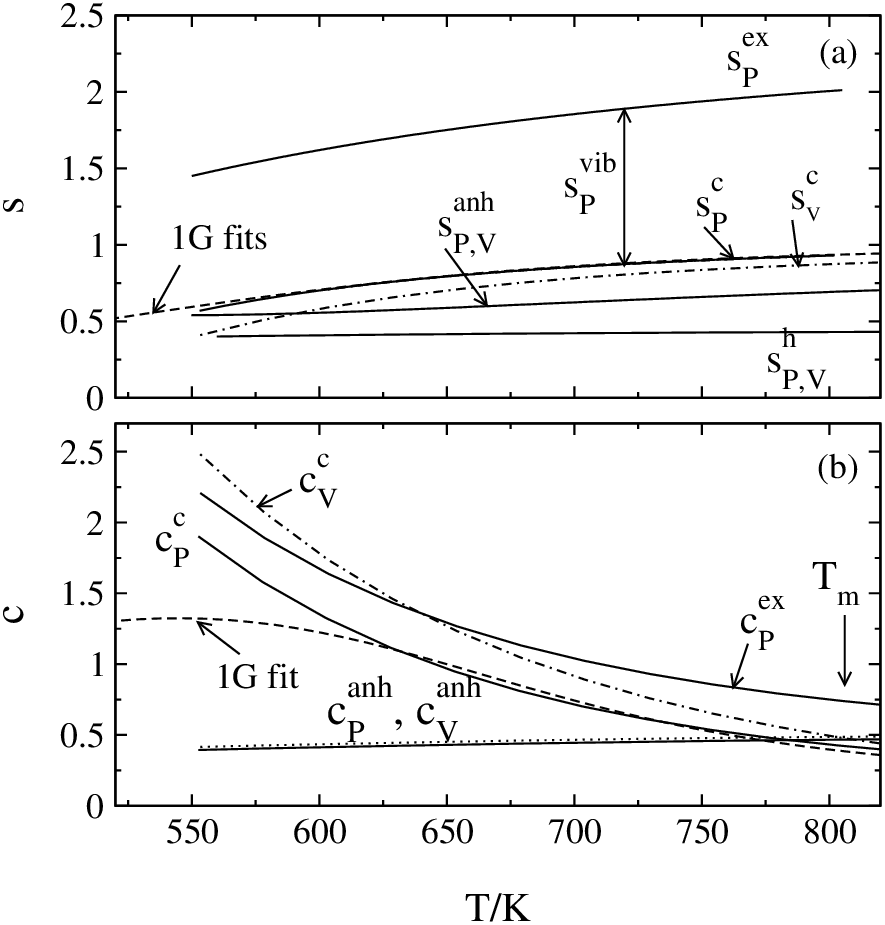}
  \caption{(a) Excess entropy of the mSW fluid over its DC crystal
    ($s^{\text{ex}}_P$) and its configurational ($s^c_{P,V}$) and
    vibrational ($s^{\text{vib}}$) components. The vibrational excess
    entropy is split into harmonic ($s^{\text{h}}_{P,V}$) and
    anharmonic ($s^{\text{anh}}_{P,V}$) parts. Configurational entropy
    at constant volume is shown by the dash-dotted line.  (b) The
    splitting of the excess heat capacity $c_P^{\text{ex}}$ into
    configurational ($c_P^c$) and anharmonic vibrational
    ($c_P^{\text{anh}}$) contributions.  The dashed lines in (a) and
    (b) refer to fits of the configurational thermodynamics to the
    excitations (1G) model.\cite{DMjcp1:07} The fit parameters $\{
    \epsilon_0,\lambda,s_0\}$ are: $\{$1675 K, 787 K, 1.0$\}$. NVT simulations were
    done at density $\rho\sigma^3=0.51$ and the density changes from 0.524 at
    $T_m$ to 0.528 in NPT simulations. }
  \label{fig:8}
\end{figure}

The splitting of the total liquid entropy into vibrational and
configurational parts is shown in the upper panel of Fig.\
\ref{fig:7}. In addition, the configurational entropy is compared in
Fig.\ \ref{fig:8} to the excess entropy over the thermodynamically
stable DC crystal in the temperature range between the melting
temperature and the lowest temperature accessible to simulations. Also
shown are the harmonic and anharmonic parts of the excess vibrational
entropy. The vibrational entropy makes about half of the overall
excess entropy, in a general accord with experimental evidence
obtained for laboratory
glass-formers.\cite{Goldstein:76,Phillips:89,Angell:04,Corezzi:04} The
situation is somewhat similar with the excess heat capacity close to
the melting point where the anharmonic vibrational part is responsible
for approximately half of the excess heat capacity (harmonic excess
heat capacity is identically zero). However, the fraction of the
anharmonic heat capacity drops down to about 15\% with lowering
temperature (Fig.\ \ref{fig:8}b).

We have used the data for the configurational thermodynamics from
simulations to fit them to the Gaussian excitations (1G)
model.\cite{DMjcp1:07} The model does not anticipate anharmonicity
playing a major role in the excess thermodynamics below the melting
temperature. Our focus is therefore limited to configurational
thermodynamics only.  As is shown in Fig.\ \ref{fig:8}, the 1G models
can be successfully fitted to the temperature dependence of the
configurational entropy and to the high-temperature portion of the
heat capacity. The model, however fails to reproduce the sharp rise of
the heat capacity at the lowest temperatures accessible to
simulations.

\subsection{Dynamics}
\begin{figure}
  \centering
  \includegraphics*[width=7cm]{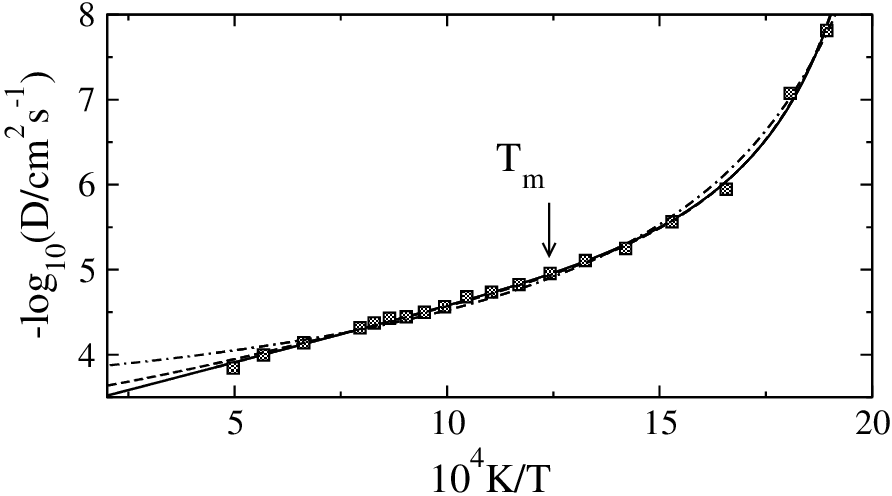}
  \caption{ Fit of the diffusivity of the mSW ($\lambda=19$) fluid from
    simulations (hatched squares) to the AG theory [dashed line, Eq.\
    (\ref{eq:2})] and to the excitations model [solid line, Eq.\
    (\ref{eq:3})]. The dashed-dotted line refers to the
    Vogel-Fulcher-Tammann (VFT) equation with the VFT temperature
    $T_0$ equal to 303 K (Table \ref{tab:1}).  The dashed and solid
    lines are indistinguishable on the scale of the plot. }
  \label{fig:9}
\end{figure}

The diffusivity data from simulations ($\lambda=19$) are shown by points in
Fig.\ \ref{fig:9}.  These results are fitted to the AG relation [Eq.\
(\ref{eq:2})] and to the dynamic equation of the Gaussian excitations
model [Eq.\ (\ref{eq:3})]. The configurational heat capacity from our
simulations is used in the fit, in contrast to the previous
application of Eq.\ (\ref{eq:3}) (Ref.\ \onlinecite{DMjcp1:07}) where
experimental dielectric relaxation data\cite{Richert:98} were fitted
to Eq.\ (\ref{eq:3}) using the excess heat capacity from the
laboratory experiment.  However, for the mSW model,
$c_P^{\text{ex}}(T)$ and $c_P^c(T)$ are off-set by almost a constant
shift of anharmonic heat capacity, and the use of either of the two to
fit diffusivity gives comparable results. The dashed line, almost
indistinguishable from the solid line in Fig.\ \ref{fig:3}, indicates
the AG relation. The Vogel-Fulcher-Tammann (VFT) equation (dash-dotted
line in Fig.\ \ref{fig:9}) gives a less satisfactory fit. In terms of
fitting the relaxation data, the AG relation is superior to both the
Gaussian excitations model and the VFT equation since it involves one
fitting parameter less.

\section{Discussion}
\begin{figure}
  \centering
  \includegraphics*[width=7cm]{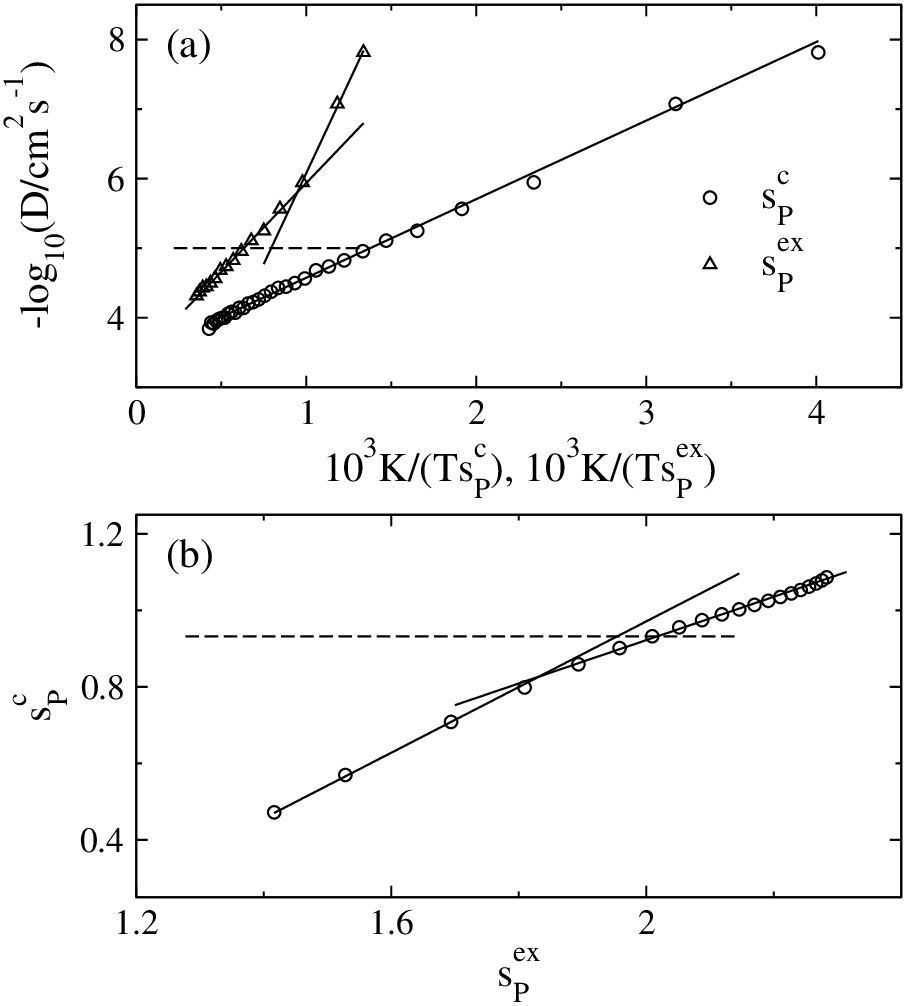}
  \caption{(a) AG plot for diffusivity using excess (triangles) and
    configurational (circles) entropy of the mSW fluid with $\lambda=19$.
    (b) $s^{\text{ex}}_P(T)$ vs $s^c_P(T)$ for the same system.  The
    dashed lines represent the values of corresponding parameters at
    the melting temperature.  }
  \label{fig:10}
\end{figure}

In discussing these results we must first recognize the sort of
frustrations that are likely to accompany any effort to resolve the
key problems of the glass transition by studying non-crystallizing
systems using MD methods. Despite the four orders of magnitude in
diffusivity that we have studied (Fig.\ \ref{fig:9}) we have barely
reached the onset of the ``low temperature domain'' (from the Stickel
temperature $T_b$ down to $T_g$) in which the Adam-Gibbs equation has
been tested experimentally using the excess entropy. It is in this
domain that experiments show linear relations between $\log D$ and
$(Ts_P^{\text{ex}})^{-1}$ predicted by the AG equation. Thus when, in
Fig.\ \ref{fig:10}, we plot our $\log D$ values against the
alternative quantities $(Ts_{P}^c)^{-1}$ and
$(Ts^{\text{ex}}_P)^{-1}$, and observe that the first is linear and
the second is not, we are not able to relate the break in the
$(Ts^{\text{ex}}_P)^{-1}$ plot to the breakdown of the AG correlation
at $T_b$ in the analysis of Richert and Angell,\cite{Richert:98} nor
to the other crossovers (Stokes-Einstein equation breakdown etc.)
observed in the experimental plots at $T_c$.\cite{Rossler:90,Mapes:06}
The break in our $\log D$ vs $(Ts^{\text{ex}}_P)^{-1}$ plot occurs at
a quite different (much higher) temperatures, where $D$ is only
$10^{-6}$ cm$^2$s$^{-1}$, and thus must have a different
origin. Whether or not there is a further break in the plot of $\log
D$ vs $(Ts^{\text{ex}}_P)^{-1}$ (or of $\log D$ vs $(Ts_{P}^c)^{-1}$
for that matter) occurring at $T_b$ cannot be told from the present
work, nor from any previous study.

This difference between the viscosity domains explored in
MD simulations of glass-formers and the low temperature domain near
$T_g$, where so many laboratory studies are carried out, is not given
adequate attention in most of the discussions of ``glassy dynamics''
simulation results.  Notwithstanding the success of well know
phenomenological models in describing the glass transition as is
observed in simulation,\cite{Giovambattista:04} the fact that one
is working above the much discussed crossover in the MD case and below
it in the experimental range of AG equation testing, cannot be
escaped. The best that can be done is to compare the slopes of the two
plots with those found in the most relevant experiments.  but little
can be gained thereby without a better theory for the AG
equation. Here we will further discuss the break in our Fig.\
\ref{fig:10} plot for diffusivities and the non-Arrhenius character of
the diffusivities, and will then seek to reconcile what seems to be a
conflict in dynamic and thermodynamic signatures of fragility in the
present system. This will provide us with the opportunity to make a
(rare) comparison of thermodynamic behavior for different potential
models of simulated glass-formers.

The most complete studies of glass-former diffusivity available are for
the cases of OTP,\cite{Mapes:06} SiO$_2$,\cite{Brebec:80} and some of
the bulk metallic glasses (BMG)\cite{Faupel:03,Bartsch:06} in which
data cover the range of diffusivity from water-like values down to
those characteristic of liquids at their glass transition temperatures
($10^{-18}$ cm$^2$s$^{-1}$).  Only in the case of OTP is the variation
of the excess entropy in the same temperature range, relative to that
of the crystal, properly known. Excess entropies relative to a mixture
of crystals, are known for some of the bulk metallic glass-formers.

The division of the excess entropy (and heat capacity) of the
supercooled liquid into vibrational and configurational components was
suggested in Goldstein's original analysis,\cite{Goldstein:76} where
it was found that in the case of OTP almost 50\% of the excess entropy
was vibrational in character. Goldstein's finding has recently been
confirmed by measurements of Wang and Richert.\cite{Richert:07}
However OTP is fragile in character and the behavior of the VDOS, in
Lewis-Wahnstr{\"o}m model, is unlike that of the present system so
comparison of our findings with those for OTP may not be
appropriate. The bulk metallic glasses, by contrast, are more similar
to the present system in their VDOS behavior (from neutron scattering
studies of their quenched and annealed states,\cite{Meyer:96} but
remember the observations were all made at fictive temperatures near
$T_g$) and prove to be relatively strong in their
kinetics.\cite{Faupel:03} Like the present system, their diffusivities
exhibit a strong Arrhenius plot curvature in the temperature range
accessible to computer simulation (thus appearing fragile in this
range) in much the same way as do classical network glasses, BeF$_2$
and SiO$_2$, at high temperatures. Thus the behavior of our mSW system
might be better compared with that of the BMG systems studied by
Chathoth et al.\cite{Chathoth:04} and reviewed by Faupel et
al.\cite{Faupel:03}

\begin{figure}
  \centering
  \includegraphics*[width=7.5cm]{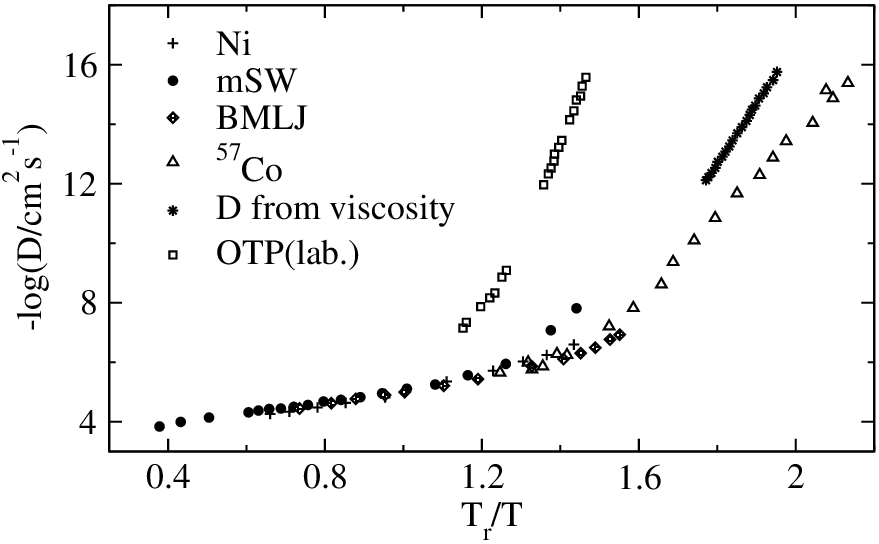}
  \caption{Diffusivity of model and laboratory glass-formers vs
    $T_r/T$, where $T_r$ is the temperature at which diffusivity is
    equal to $10^{-5}$ cm$^2$s$^{-1}$ (the end of the Arrhenius
    region).  Model fluids: larger component of the BMLJ
    liquid\cite{Sastry:01} and mSW ($\lambda=19$) fluid. Laboratory liquids:
    $^{57}$Co (triangles)\cite{Bartsch:06} and Ni
    (pluses)\cite{Chathoth:04} tracers in
    Pd$_{43}$Cu$_{27}$Ni$_{10}$P$_{20}$ melt and OTP
    (squares).\cite{Mapes:06} Stars refer to the diffusivity of
    $^{57}$Co calculated from the Stokes-Einstein
    equation.\cite{Bartsch:06} }
  \label{fig:11}
\end{figure}

\begin{figure}
  \centering

  \includegraphics*[width=7cm]{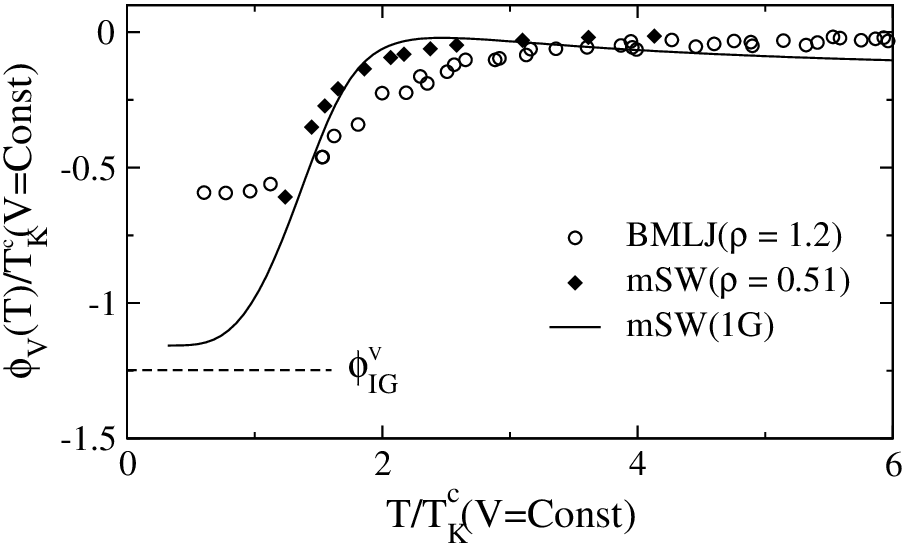}
  \caption{Excitation profiles ($\bar\phi_V(T)$) of the mSW liquid
    (diamonds, $\lambda=19$) vs temperature measured on the Kauzmann
    temperature scale; $T_K^c(V=Const)$ is from extrapolating
    $s_V^c(T)$ to zero.  Solid line is the excitation profile obtained
    from the fit of 1G model to the constant-volume configurational
    entropy of the mSW liquid obtained by thermodynamic integration
    from the ideal gas reference state, and horizontal dashed line
    marks the position of the ideal glass basin depth.  Open circles
    are data from the slowest effective cooling rate set of inherent
    structure energies from the study of Sastry et al on BMLJ at $\rho\sigma^3
    =1.213$\cite{Sastry:98} scaled using the Kauzmann temperature
    $T_K^c=34.35$.\cite{Sastry:01} The level at low T in this case is
    not the ground state but the glassy state frozen in at this
    cooling rate. The sharper profile for the mSW model implies a more
    fragile liquid than the BMLJ model, as is also seen in the
    diffusivity data in Fig.\ \ref{fig:11}. }
  \label{fig:12}
\end{figure}

In Fig.\ \ref{fig:11}, we compare the diffusivities from Fig.\
\ref{fig:9} with those of various components of the bulk glass-former
Pd$_{43}$Ni$_{10}$Cu$_{13}$P$_{20}$ from Refs.\
\onlinecite{Chathoth:04} and \onlinecite{Bartsch:06} after scaling by
the temperature at which each system exhibits $D = 10^{-5}$
cm$^2$s$^{-1}$ (near where the deviation from Arrhenius behavior first
becomes obvious). On the larger temperature scale the Ni diffusivity,
like the $^{57}$Co diffusivity of Ref.\ \onlinecite{Bartsch:06} and
the viscosity of Ref.\ \onlinecite{Bartsch:06}, all return to
Arrhenius behavior with a larger slope, and the behavior appears
non-fragile, approximately like glycerol. Thus the strong curvature
which lead Molinero et al.\cite{Molinero:06} to conclude that mSW is
very fragile, does not necessarily continue. This would rationalize
what otherwise is a problem raised by the excitation profile of Fig.\
\ref{fig:4} - which is not of the form expected for a very fragile
liquid according to the 1G model. This point is illustrated in Fig.\
\ref{fig:12} where the excitation profiles of the BMLJ and mSW liquids
are scaled onto the same plot by the use of their Kauzmann
temperatures ($T_K^c=34.35$ K\cite{Sastry:01} for BMLJ and $T_K=434$ K
for mSW, Table \ref{tab:1}). Unlike the more fragile
cases,\cite{DMjcp5:05,DMjcp1:07} which develop an S-shaped profile and
exhibit phase transitions (not unlike that of silicon itself), these
profiles always have positive slopes.  Consistent with the difference
in their diffusivity behavior seen in Fig.\ \ref{fig:11}, the profile
for mSW is sharper than that of the less fragile BMLJ. Figures
\ref{fig:11} and \ref{fig:12} together provide the best evidence to
date of the surprisingly non-fragile behavior of the much studied BMLJ
system.

\begin{figure}
  \centering
  \includegraphics*[width=7cm]{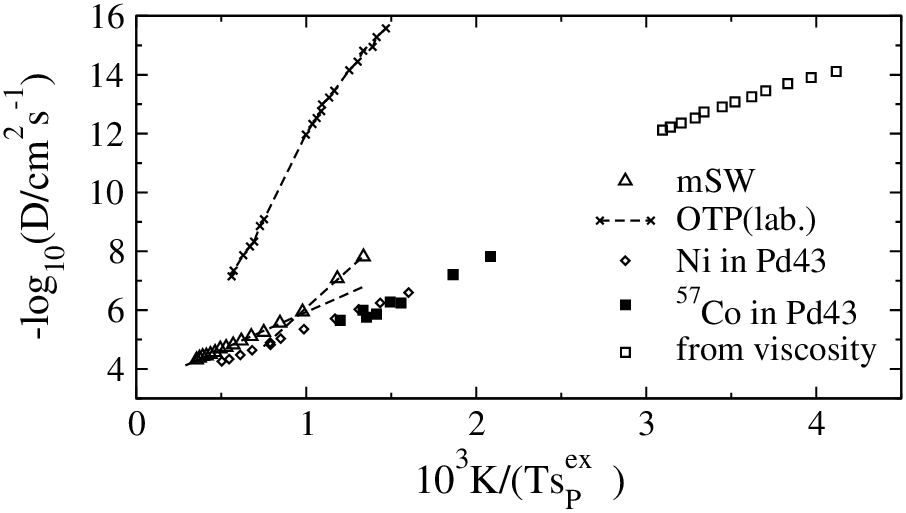} 
  \caption{Comparison of AG plots of laboratory glass-formers to the
    simulation results of mSW ($\lambda=19$).  Laboratory glass-formers: OTP
    (crosses\cite{Mapes:06}); Ni (diamonds\cite{Chathoth:04}) and
    $^{57}$Co (closed squares\cite{Bartsch:06}) tracers in
    Pd$_{43}$Cu$_{27}$Ni$_{10}$P$_{20}$ melt.  Open squares refer to
    the diffusivity of the same melt calculated from viscosity by
    using the Stokes-Einstein equation.\cite{Bartsch:06}}
  \label{fig:13}
\end{figure}

A way of inducing fragile character in an atomic system spherically
symmetric potential is, according to Sastry, to increase the density
of BMLJ. This was shown to increase the slope of the AG plot, Fig.\
\ref{fig:10}a, in the same way that is seen when the experimental
data for OTP are added to the plot. We demonstrate this in Fig.\
\ref{fig:13}. To show consistency, we include, in Fig.\ \ref{fig:13},
the data for bulk metallic glasses, using the excess entropy data of
the BMG reported by Kuno et al.\cite{Kuno:04} Although these data
refer to the melting of ternary eutectic composition, we consider that
the fusion enthalpy used in the entropy assessment to be valid
(i.e. to contain negligible non-ideal mixing enthalpy), because it has
been shown that the crystals that fuse at the eutectic are already
binary compounds. The slope of the plot for the BMG diffusivities for
Ni and Co, which are decoupled from the viscosity, is less than for
mSW using excess entropy, but the slope for the viscosity-based data
of Fig.\ \ref{fig:11} is essentially the same.  The variations in
slope in Fig. 13, however, are not well accounted for. Here we recall
that, in Ref.\ \onlinecite{DMjcp1:07}, such plots could be reduced to
an all common slope using Eq.\ (\ref{eq:3}) of this paper.

\begin{figure}
  \centering
  \includegraphics*[width=7cm]{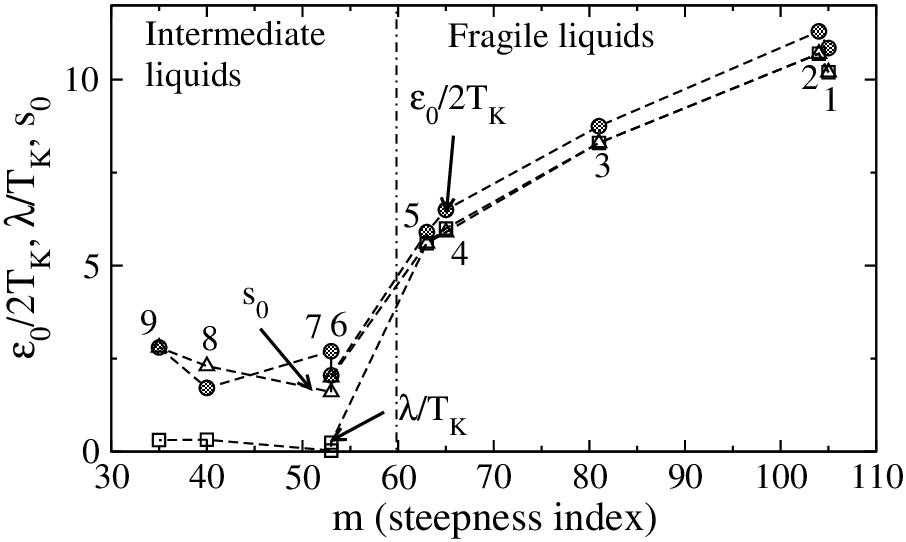}
  \caption{Excitation energy $\epsilon_0$, the trapping energy parameter $\lambda$,
    and the excitation entropy $s_0$ of some laboratory glass-formers.
    The energy parameters are scaled with the experimental Kauzmann
    temperature $T_K$. The parameters obtained by fitting the
    excitation 1G model to experimental excess thermodynamic data are
    plotted against the steepness fragility parameter
    $m$.\cite{Angell:95} The fragility number $m\simeq 60$ separates
    fragile from intermediate liquids (dash-dotted line).  The numbers
    in the plot indicate: toluene (1), D,L-propene carbonate (2), OTP
    (3), 2-methyltetrahydrofurane (4), salol (5), 3-bromopentane (6),
    glycerol (7), selenium (8), and $n$-propanol (9).  For selenim,
    the steepness index of $m=40$ calculated in Ref.\
    \onlinecite{Richert:06} 
    was used. }
  \label{fig:14}
\end{figure}

The analysis of the thermodynamic and relaxation data of laboratory
glass-formers has allowed us to classify those liquids according to
thermodynamic parameters of their configurational
excitations.\cite{DMjcp5:05,DMjcp1:07} The excitations model describes
the thermodynamics of a supercooled liquid as an ideal gas of
excitations each carrying excitation energy $\epsilon$ and entropy $s_0$.
The energy of excitations belongs to a Gaussian manifold with the
average excitation energy $\epsilon_0-2x(T)\lambda$ ($x(T)$ is the population of
the excited state) and the variance $2T\lambda$. This representation is in
fact equivalent to an ensemble of excitations with mean-field
attractions.  When the laboratory data for fragile liquids are fitted
to the 1G model, the excitation parameters show universality when the
energies are scaled with the Kauzmann temperature. In these reduced
units, the excitation entropy $s_0$ becomes the only relevant
parameter determining fragility as shown in Fig.\ \ref{fig:14} where
the data are plotted against the steepness fragility
index:\cite{Richert:06}
\begin{equation}
  \label{eq:23}
  \frac{\epsilon_0}{2T_K} \simeq \frac{\lambda}{T_K} \simeq s_0 .
\end{equation}
This universality does not hold for liquids of intermediate fragility
(intermediate liquids). Noteworthy is a much smaller width parameter
$\lambda$ for intermediate liquids compared to fragile liquids (Fig.\
\ref{fig:14}).

The current simulations of the mSW liquid generally support the basic
assumptions incorporated in the development of the excitations model
of low-temperature glass-formers,\cite{DMjcp5:05,DMjcp1:07} although,
as for the comparison to laboratory data, the MD simulations probe the
temperature range much higher than the one anticipated in the theory
development ($T_g\leq T \leq T_b$). Nevertheless, some conclusions can be
drawn. The 1G model neglects the temperature dependence of the
excitation entropy and thus the anharmonicity effects. Anharmonic
vibrations are significant in the potential energy landscape of
the mSW model at high temperatures, but their contribution to the
excess heat capacity drops to about $\simeq 15$ \% in the lowest portion of
the temperature range studied here. Therefore, the assumption of the
temperature-independent entropy of elementary excitations might be a
good first-order approximation in applications of the model to data at
low temperatures close to $T_g$.

The most significant ingredient of the excitations model requiring
testing on available data is the anticipated temperature dependence of
the effective Gaussian width of the distribution of basin
energies. The temperature-dependent width is introduced into the model
to account for an effectively non-Gaussian landscape probed by the
system when exploring deeper basins with lowering temperature. An
exact non-Gaussian landscape model recently developed by one of
us\cite{DMpre:07} has taught us that this temperature dependence can
be quite complex such that a linear scaling probably applies only to a
limited range of temperatures. Even though, in the range of
temperatures studied by our simulations, the effective width scales
linearly with temperature for both constant-volume and
constant-pressure ensembles (Fig.\ \ref{fig:4}).  Unfortunately, the
present simulations do not allow access to the lower portion of the
landscape, and deviations from gaussianity observed in some recent
simulations\cite{Moreno:06,DMpre:07} are not pronounced in the
enumeration function $\omega(\phi)$.

\begin{acknowledgments}
  This work was supported by the NSF through the grants CHE-0616646
  (V.\ K.\ and D.\ V.\ M.) and DMR-0082535 (C.\ A.\ A.). We are
  grateful to Srikanth Sastry for sharing his experience with
  numerical calculations.
\end{acknowledgments}  

\bibliographystyle{apsrev}
\bibliography{/home/dmitry/p/bib/chem_abbr,/home/dmitry/p/bib/photosynth,/home/dmitry/p/bib/liquids,/home/dmitry/p/bib/glass,/home/dmitry/p/bib/et,/home/dmitry/p/bib/dm,/home/dmitry/p/bib/dynamics,/home/dmitry/p/bib/ferro}

\begin{thebibliography}{69}
\expandafter\ifx\csname natexlab\endcsname\relax\def\natexlab#1{#1}\fi
\expandafter\ifx\csname bibnamefont\endcsname\relax
  \def\bibnamefont#1{#1}\fi
\expandafter\ifx\csname bibfnamefont\endcsname\relax
  \def\bibfnamefont#1{#1}\fi
\expandafter\ifx\csname citenamefont\endcsname\relax
  \def\citenamefont#1{#1}\fi
\expandafter\ifx\csname url\endcsname\relax
  \def\url#1{\texttt{#1}}\fi
\expandafter\ifx\csname urlprefix\endcsname\relax\def\urlprefix{URL }\fi
\providecommand{\bibinfo}[2]{#2}
\providecommand{\eprint}[2][]{\url{#2}}

\bibitem[{\citenamefont{Ngai}(2000)}]{Ngai:00}
\bibinfo{author}{\bibfnamefont{K.~L.} \bibnamefont{Ngai}}, \bibinfo{journal}{J.
  Non-Cryst. Sol.} \textbf{\bibinfo{volume}{275}}, \bibinfo{pages}{7}
  (\bibinfo{year}{2000}).

\bibitem[{\citenamefont{Angell}(1995)}]{Angell:95}
\bibinfo{author}{\bibfnamefont{C.~A.} \bibnamefont{Angell}},
  \bibinfo{journal}{Science} \textbf{\bibinfo{volume}{267}},
  \bibinfo{pages}{1924} (\bibinfo{year}{1995}).

\bibitem[{\citenamefont{Goldstein}(1969)}]{Goldstein:69}
\bibinfo{author}{\bibfnamefont{M.}~\bibnamefont{Goldstein}},
  \bibinfo{journal}{J. Chem. Phys.} \textbf{\bibinfo{volume}{51}},
  \bibinfo{pages}{3728} (\bibinfo{year}{1969}).

\bibitem[{\citenamefont{Adam and Gibbs}(1965)}]{Adam:65}
\bibinfo{author}{\bibfnamefont{G.}~\bibnamefont{Adam}} \bibnamefont{and}
  \bibinfo{author}{\bibfnamefont{J.~H.} \bibnamefont{Gibbs}},
  \bibinfo{journal}{J. Chem. Phys.} \textbf{\bibinfo{volume}{43}},
  \bibinfo{pages}{139} (\bibinfo{year}{1965}).

\bibitem[{\citenamefont{Xia and Wolynes}(2000)}]{Xia:00}
\bibinfo{author}{\bibfnamefont{X.}~\bibnamefont{Xia}} \bibnamefont{and}
  \bibinfo{author}{\bibfnamefont{P.~G.} \bibnamefont{Wolynes}},
  \bibinfo{journal}{Proc. Nat. Acad. Sci.} \textbf{\bibinfo{volume}{97}},
  \bibinfo{pages}{2990} (\bibinfo{year}{2000}).

\bibitem[{\citenamefont{Xia and Wolynes}(2001)}]{Xia:01}
\bibinfo{author}{\bibfnamefont{X.}~\bibnamefont{Xia}} \bibnamefont{and}
  \bibinfo{author}{\bibfnamefont{P.~G.} \bibnamefont{Wolynes}},
  \bibinfo{journal}{J. Phys. Chem. B} \textbf{\bibinfo{volume}{105}},
  \bibinfo{pages}{6570} (\bibinfo{year}{2001}).

\bibitem[{\citenamefont{Lubchenko and Wolynes}(2004)}]{Lubchenko:04}
\bibinfo{author}{\bibfnamefont{V.}~\bibnamefont{Lubchenko}} \bibnamefont{and}
  \bibinfo{author}{\bibfnamefont{P.~G.} \bibnamefont{Wolynes}},
  \bibinfo{journal}{J. Chem. Phys.} \textbf{\bibinfo{volume}{121}},
  \bibinfo{pages}{2852} (\bibinfo{year}{2004}).

\bibitem[{\citenamefont{B{\"a}ssler}(1987)}]{Baessler:87}
\bibinfo{author}{\bibfnamefont{H.}~\bibnamefont{B{\"a}ssler}},
  \bibinfo{journal}{Phys. Rev. Lett.} \textbf{\bibinfo{volume}{58}},
  \bibinfo{pages}{767} (\bibinfo{year}{1987}).

\bibitem[{\citenamefont{Arhipov and B{\"a}ssler}(1994)}]{Arhipov:94}
\bibinfo{author}{\bibfnamefont{V.~I.} \bibnamefont{Arhipov}} \bibnamefont{and}
  \bibinfo{author}{\bibfnamefont{H.}~\bibnamefont{B{\"a}ssler}},
  \bibinfo{journal}{J. Phys. Chem.} \textbf{\bibinfo{volume}{98}},
  \bibinfo{pages}{662} (\bibinfo{year}{1994}).

\bibitem[{\citenamefont{Dyre}(1995)}]{Dyre:95}
\bibinfo{author}{\bibfnamefont{J.~C.} \bibnamefont{Dyre}},
  \bibinfo{journal}{Phys. Rev. B} \textbf{\bibinfo{volume}{51}},
  \bibinfo{pages}{12276} (\bibinfo{year}{1995}).

\bibitem[{\citenamefont{Matyushov and Angell}(2005)}]{DMjcp5:05}
\bibinfo{author}{\bibfnamefont{D.~V.} \bibnamefont{Matyushov}}
  \bibnamefont{and} \bibinfo{author}{\bibfnamefont{C.~A.}
  \bibnamefont{Angell}}, \bibinfo{journal}{J. Chem. Phys.}
  \textbf{\bibinfo{volume}{123}}, \bibinfo{pages}{034506}
  (\bibinfo{year}{2005}).

\bibitem[{\citenamefont{Dalle-Ferrier et~al.}(2007)\citenamefont{Dalle-Ferrier,
  Thibierge, Abla-Simionesco, Berthier, Biroli, Bouchaud, Ladieu, L'H{\^o}te,
  and Tarjus}}]{Dalle-Ferrier:07}
\bibinfo{author}{\bibfnamefont{C.}~\bibnamefont{Dalle-Ferrier}},
  \bibinfo{author}{\bibfnamefont{C.}~\bibnamefont{Thibierge}},
  \bibinfo{author}{\bibfnamefont{C.}~\bibnamefont{Abla-Simionesco}},
  \bibinfo{author}{\bibfnamefont{L.}~\bibnamefont{Berthier}},
  \bibinfo{author}{\bibfnamefont{G.}~\bibnamefont{Biroli}},
  \bibinfo{author}{\bibfnamefont{J.-P.} \bibnamefont{Bouchaud}},
  \bibinfo{author}{\bibfnamefont{F.}~\bibnamefont{Ladieu}},
  \bibinfo{author}{\bibfnamefont{D.}~\bibnamefont{L'H{\^o}te}},
  \bibnamefont{and} \bibinfo{author}{\bibfnamefont{G.}~\bibnamefont{Tarjus}},
  \bibinfo{journal}{Phys. Rev. E} \textbf{\bibinfo{volume}{76}},
  \bibinfo{pages}{041510} (\bibinfo{year}{2007}).

\bibitem[{\citenamefont{Sastry}(2001)}]{Sastry:01}
\bibinfo{author}{\bibfnamefont{S.}~\bibnamefont{Sastry}},
  \bibinfo{journal}{Nature} \textbf{\bibinfo{volume}{409}},
  \bibinfo{pages}{164} (\bibinfo{year}{2001}).

\bibitem[{\citenamefont{Saika-Voivod et~al.}(2001)\citenamefont{Saika-Voivod,
  Poole, and Sciortino}}]{Voivod:01}
\bibinfo{author}{\bibfnamefont{I.}~\bibnamefont{Saika-Voivod}},
  \bibinfo{author}{\bibfnamefont{P.~H.} \bibnamefont{Poole}}, \bibnamefont{and}
  \bibinfo{author}{\bibfnamefont{F.}~\bibnamefont{Sciortino}},
  \bibinfo{journal}{Nature} \textbf{\bibinfo{volume}{412}},
  \bibinfo{pages}{514} (\bibinfo{year}{2001}).

\bibitem[{\citenamefont{Mossa et~al.}(2002)\citenamefont{Mossa, LaNave,
  Stanley, Donati, Sciortino, and Tartaglia}}]{Mossa:02}
\bibinfo{author}{\bibfnamefont{S.}~\bibnamefont{Mossa}},
  \bibinfo{author}{\bibfnamefont{E.}~\bibnamefont{LaNave}},
  \bibinfo{author}{\bibfnamefont{H.~E.} \bibnamefont{Stanley}},
  \bibinfo{author}{\bibfnamefont{C.}~\bibnamefont{Donati}},
  \bibinfo{author}{\bibfnamefont{F.}~\bibnamefont{Sciortino}},
  \bibnamefont{and}
  \bibinfo{author}{\bibfnamefont{P.}~\bibnamefont{Tartaglia}},
  \bibinfo{journal}{Phys. Rev. E} \textbf{\bibinfo{volume}{65}},
  \bibinfo{pages}{041205} (\bibinfo{year}{2002}).

\bibitem[{\citenamefont{Giovambattista
  et~al.}(2003)\citenamefont{Giovambattista, Buldyrev, Starr, and
  Stanley}}]{Giovambattista:03}
\bibinfo{author}{\bibfnamefont{N.}~\bibnamefont{Giovambattista}},
  \bibinfo{author}{\bibfnamefont{S.~V.} \bibnamefont{Buldyrev}},
  \bibinfo{author}{\bibfnamefont{F.~W.} \bibnamefont{Starr}}, \bibnamefont{and}
  \bibinfo{author}{\bibfnamefont{H.~E.} \bibnamefont{Stanley}},
  \bibinfo{journal}{Phys. Rev. Lett.} \textbf{\bibinfo{volume}{90}},
  \bibinfo{pages}{085506} (\bibinfo{year}{2003}).

\bibitem[{\citenamefont{Saika-Voivod et~al.}(2004)\citenamefont{Saika-Voivod,
  Sciortino, and Poole}}]{Voivod:04}
\bibinfo{author}{\bibfnamefont{I.}~\bibnamefont{Saika-Voivod}},
  \bibinfo{author}{\bibfnamefont{F.}~\bibnamefont{Sciortino}},
  \bibnamefont{and} \bibinfo{author}{\bibfnamefont{P.~H.} \bibnamefont{Poole}},
  \bibinfo{journal}{Phys. Rev. E} \textbf{\bibinfo{volume}{69}},
  \bibinfo{pages}{041503} (\bibinfo{year}{2004}).

\bibitem[{\citenamefont{Gebremichael et~al.}(2005)\citenamefont{Gebremichael,
  Vogel, Bergroth, Starr, and Glotzer}}]{Gebremichael:05}
\bibinfo{author}{\bibfnamefont{Y.}~\bibnamefont{Gebremichael}},
  \bibinfo{author}{\bibfnamefont{M.}~\bibnamefont{Vogel}},
  \bibinfo{author}{\bibfnamefont{M.~N.~J.} \bibnamefont{Bergroth}},
  \bibinfo{author}{\bibfnamefont{F.~W.} \bibnamefont{Starr}}, \bibnamefont{and}
  \bibinfo{author}{\bibfnamefont{S.~C.} \bibnamefont{Glotzer}},
  \bibinfo{journal}{J. Phys. Chem. B} \textbf{\bibinfo{volume}{109}},
  \bibinfo{pages}{15068} (\bibinfo{year}{2005}).

\bibitem[{\citenamefont{Angell}(1991)}]{Angell:91}
\bibinfo{author}{\bibfnamefont{C.~A.} \bibnamefont{Angell}},
  \bibinfo{journal}{J. Non-Cryst. Solids} \textbf{\bibinfo{volume}{131-133}},
  \bibinfo{pages}{13} (\bibinfo{year}{1991}).

\bibitem[{\citenamefont{Richert and Angell}(1998)}]{Richert:98}
\bibinfo{author}{\bibfnamefont{R.}~\bibnamefont{Richert}} \bibnamefont{and}
  \bibinfo{author}{\bibfnamefont{A.~C.} \bibnamefont{Angell}},
  \bibinfo{journal}{J. Chem. Phys.} \textbf{\bibinfo{volume}{108}},
  \bibinfo{pages}{9016} (\bibinfo{year}{1998}).

\bibitem[{\citenamefont{Stillinger and Weber}(1982)}]{Stillinger:82}
\bibinfo{author}{\bibfnamefont{F.~H.} \bibnamefont{Stillinger}}
  \bibnamefont{and} \bibinfo{author}{\bibfnamefont{T.~A.} \bibnamefont{Weber}},
  \bibinfo{journal}{Phys. Rev. A} \textbf{\bibinfo{volume}{25}},
  \bibinfo{pages}{978} (\bibinfo{year}{1982}).

\bibitem[{\citenamefont{Goldstein}(1976)}]{Goldstein:76}
\bibinfo{author}{\bibfnamefont{M.~J.} \bibnamefont{Goldstein}},
  \bibinfo{journal}{J. Chem. Phys.} \textbf{\bibinfo{volume}{64}},
  \bibinfo{pages}{4767} (\bibinfo{year}{1976}).

\bibitem[{\citenamefont{Phillips et~al.}(1989)\citenamefont{Phillips, Buchenau,
  N{\"u}cker, Dianoux, and Petry}}]{Phillips:89}
\bibinfo{author}{\bibfnamefont{W.~A.} \bibnamefont{Phillips}},
  \bibinfo{author}{\bibfnamefont{U.}~\bibnamefont{Buchenau}},
  \bibinfo{author}{\bibfnamefont{N.}~\bibnamefont{N{\"u}cker}},
  \bibinfo{author}{\bibfnamefont{A.-J.} \bibnamefont{Dianoux}},
  \bibnamefont{and} \bibinfo{author}{\bibfnamefont{W.}~\bibnamefont{Petry}},
  \bibinfo{journal}{Phys. Rev. Lett.} \textbf{\bibinfo{volume}{63}},
  \bibinfo{pages}{2381} (\bibinfo{year}{1989}).

\bibitem[{\citenamefont{Corezzi et~al.}(2004)\citenamefont{Corezzi, Comez, and
  Fioretto}}]{Corezzi:04}
\bibinfo{author}{\bibfnamefont{S.}~\bibnamefont{Corezzi}},
  \bibinfo{author}{\bibfnamefont{L.}~\bibnamefont{Comez}}, \bibnamefont{and}
  \bibinfo{author}{\bibfnamefont{D.}~\bibnamefont{Fioretto}},
  \bibinfo{journal}{Eur. Phys. J. E} \textbf{\bibinfo{volume}{14}},
  \bibinfo{pages}{143} (\bibinfo{year}{2004}).

\bibitem[{\citenamefont{Martinez and Angell}(2001)}]{Martinez:01}
\bibinfo{author}{\bibfnamefont{L.-M.} \bibnamefont{Martinez}} \bibnamefont{and}
  \bibinfo{author}{\bibfnamefont{C.~A.} \bibnamefont{Angell}},
  \bibinfo{journal}{Nature} \textbf{\bibinfo{volume}{410}},
  \bibinfo{pages}{663} (\bibinfo{year}{2001}).

\bibitem[{\citenamefont{Angell and Borick}(2002)}]{Angell:02}
\bibinfo{author}{\bibfnamefont{C.~A.} \bibnamefont{Angell}} \bibnamefont{and}
  \bibinfo{author}{\bibfnamefont{S.}~\bibnamefont{Borick}},
  \bibinfo{journal}{J. Non-Crystal. Solids} \textbf{\bibinfo{volume}{307-310}},
  \bibinfo{pages}{393} (\bibinfo{year}{2002}).

\bibitem[{\citenamefont{Starr et~al.}(2003)\citenamefont{Starr, Angell, and
  Stanley}}]{Starr:03}
\bibinfo{author}{\bibfnamefont{F.~W.} \bibnamefont{Starr}},
  \bibinfo{author}{\bibfnamefont{C.~A.} \bibnamefont{Angell}},
  \bibnamefont{and} \bibinfo{author}{\bibfnamefont{H.~E.}
  \bibnamefont{Stanley}}, \bibinfo{journal}{Physica A}
  \textbf{\bibinfo{volume}{323}}, \bibinfo{pages}{51} (\bibinfo{year}{2003}).

\bibitem[{\citenamefont{Johari}(2007)}]{Johari:07}
\bibinfo{author}{\bibfnamefont{G.~P.} \bibnamefont{Johari}},
  \bibinfo{journal}{J. Chem. Phys.} \textbf{\bibinfo{volume}{126}},
  \bibinfo{eid}{114901} (\bibinfo{year}{2007}).

\bibitem[{\citenamefont{Lu et~al.}(2002)\citenamefont{Lu, G{\"o}rler, and
  Willnecker}}]{Lu:02}
\bibinfo{author}{\bibfnamefont{I.-R.} \bibnamefont{Lu}},
  \bibinfo{author}{\bibfnamefont{G.~P.} \bibnamefont{G{\"o}rler}},
  \bibnamefont{and}
  \bibinfo{author}{\bibfnamefont{R.}~\bibnamefont{Willnecker}},
  \bibinfo{journal}{Appl. Phys. Lett.} \textbf{\bibinfo{volume}{80}},
  \bibinfo{pages}{4534} (\bibinfo{year}{2002}).

\bibitem[{\citenamefont{Moynihan and Angell}(2000)}]{Moynihan:00}
\bibinfo{author}{\bibfnamefont{C.~T.} \bibnamefont{Moynihan}} \bibnamefont{and}
  \bibinfo{author}{\bibfnamefont{C.~A.} \bibnamefont{Angell}},
  \bibinfo{journal}{J. Non-Crystal. Sol.} \textbf{\bibinfo{volume}{274}},
  \bibinfo{pages}{131} (\bibinfo{year}{2000}).

\bibitem[{\citenamefont{Kob and Andersen}(1994)}]{KobAndersen:94}
\bibinfo{author}{\bibfnamefont{W.}~\bibnamefont{Kob}} \bibnamefont{and}
  \bibinfo{author}{\bibfnamefont{H.~C.} \bibnamefont{Andersen}},
  \bibinfo{journal}{Phys. Rev. Lett.} \textbf{\bibinfo{volume}{73}},
  \bibinfo{pages}{1376} (\bibinfo{year}{1994}).

\bibitem[{\citenamefont{Matyushov and Angell}(2007)}]{DMjcp1:07}
\bibinfo{author}{\bibfnamefont{D.~V.} \bibnamefont{Matyushov}}
  \bibnamefont{and} \bibinfo{author}{\bibfnamefont{C.~A.}
  \bibnamefont{Angell}}, \bibinfo{journal}{J. Chem. Phys.}
  \textbf{\bibinfo{volume}{126}}, \bibinfo{pages}{094501}
  (\bibinfo{year}{2007}).

\bibitem[{\citenamefont{Angell et~al.}(2003)\citenamefont{Angell, Yue, Wang,
  Copley, Borick, and Mossa}}]{Angell:03}
\bibinfo{author}{\bibfnamefont{C.~A.} \bibnamefont{Angell}},
  \bibinfo{author}{\bibfnamefont{Y.}~\bibnamefont{Yue}},
  \bibinfo{author}{\bibfnamefont{L.-M.} \bibnamefont{Wang}},
  \bibinfo{author}{\bibfnamefont{J.~R.~D.} \bibnamefont{Copley}},
  \bibinfo{author}{\bibfnamefont{S.}~\bibnamefont{Borick}}, \bibnamefont{and}
  \bibinfo{author}{\bibfnamefont{S.}~\bibnamefont{Mossa}}, \bibinfo{journal}{J.
  Phys.: Condens. Matter} \textbf{\bibinfo{volume}{15}}, \bibinfo{pages}{S1051}
  (\bibinfo{year}{2003}).

\bibitem[{\citenamefont{Wang and Richert}(2007)}]{Richert:07}
\bibinfo{author}{\bibfnamefont{L.-M.} \bibnamefont{Wang}} \bibnamefont{and}
  \bibinfo{author}{\bibfnamefont{R.}~\bibnamefont{Richert}},
  \bibinfo{journal}{Phys. Rev. Lett.} \textbf{\bibinfo{volume}{99}},
  \bibinfo{pages}{185701} (\bibinfo{year}{2007}).

\bibitem[{\citenamefont{Lewis and Wahnstr\"om}(1994)}]{LewisW:94}
\bibinfo{author}{\bibfnamefont{L.~J.} \bibnamefont{Lewis}} \bibnamefont{and}
  \bibinfo{author}{\bibfnamefont{G.}~\bibnamefont{Wahnstr\"om}},
  \bibinfo{journal}{Phys. Rev. E} \textbf{\bibinfo{volume}{50}},
  \bibinfo{pages}{3865} (\bibinfo{year}{1994}).

\bibitem[{\citenamefont{Molinero et~al.}(2006)\citenamefont{Molinero, Sastry,
  and Angell}}]{Molinero:06}
\bibinfo{author}{\bibfnamefont{V.}~\bibnamefont{Molinero}},
  \bibinfo{author}{\bibfnamefont{S.}~\bibnamefont{Sastry}}, \bibnamefont{and}
  \bibinfo{author}{\bibfnamefont{C.~A.} \bibnamefont{Angell}},
  \bibinfo{journal}{Phys. Rev. Lett.} \textbf{\bibinfo{volume}{97}},
  \bibinfo{eid}{075701} (\bibinfo{year}{2006}).

\bibitem[{\citenamefont{Scala et~al.}(2000)\citenamefont{Scala, Starr, LaNave,
  Sciortino, and Stanley}}]{Scala:00}
\bibinfo{author}{\bibfnamefont{A.}~\bibnamefont{Scala}},
  \bibinfo{author}{\bibfnamefont{F.~W.} \bibnamefont{Starr}},
  \bibinfo{author}{\bibfnamefont{E.}~\bibnamefont{LaNave}},
  \bibinfo{author}{\bibfnamefont{F.}~\bibnamefont{Sciortino}},
  \bibnamefont{and} \bibinfo{author}{\bibfnamefont{H.~E.}
  \bibnamefont{Stanley}}, \bibinfo{journal}{Nature}
  \textbf{\bibinfo{volume}{406}}, \bibinfo{pages}{166} (\bibinfo{year}{2000}).

\bibitem[{\citenamefont{Ghorai and Matyushov}(2006)}]{DMjpcb1:06}
\bibinfo{author}{\bibfnamefont{P.~K.} \bibnamefont{Ghorai}} \bibnamefont{and}
  \bibinfo{author}{\bibfnamefont{D.~V.} \bibnamefont{Matyushov}},
  \bibinfo{journal}{J. Phys. Chem. B} \textbf{\bibinfo{volume}{110}},
  \bibinfo{pages}{1866} (\bibinfo{year}{2006}).

\bibitem[{\citenamefont{Xu et~al.}(2005)\citenamefont{Xu, Kumar, Buldyrev,
  Chen, Poole, Sciortino, and Stanley}}]{Xu:05}
\bibinfo{author}{\bibfnamefont{L.}~\bibnamefont{Xu}},
  \bibinfo{author}{\bibfnamefont{P.}~\bibnamefont{Kumar}},
  \bibinfo{author}{\bibfnamefont{S.~V.} \bibnamefont{Buldyrev}},
  \bibinfo{author}{\bibfnamefont{S.-H.} \bibnamefont{Chen}},
  \bibinfo{author}{\bibfnamefont{P.~H.} \bibnamefont{Poole}},
  \bibinfo{author}{\bibfnamefont{F.}~\bibnamefont{Sciortino}},
  \bibnamefont{and} \bibinfo{author}{\bibfnamefont{H.~E.}
  \bibnamefont{Stanley}}, \bibinfo{journal}{Proc. Natl. Acad. Sci.}
  \textbf{\bibinfo{volume}{102}}, \bibinfo{pages}{16558}
  (\bibinfo{year}{2005}).

\bibitem[{\citenamefont{Angell and Wong}(1970)}]{Angell:70}
\bibinfo{author}{\bibfnamefont{C.~A.} \bibnamefont{Angell}} \bibnamefont{and}
  \bibinfo{author}{\bibfnamefont{J.}~\bibnamefont{Wong}}, \bibinfo{journal}{J.
  Chem. Phys.} \textbf{\bibinfo{volume}{53}}, \bibinfo{pages}{2053}
  (\bibinfo{year}{1970}).

\bibitem[{\citenamefont{Stillinger}(1998)}]{StillingerJPCB:98}
\bibinfo{author}{\bibfnamefont{F.}~\bibnamefont{Stillinger}},
  \bibinfo{journal}{J.\ Phys.\ Chem.\ B} \textbf{\bibinfo{volume}{102}},
  \bibinfo{pages}{2807} (\bibinfo{year}{1998}).

\bibitem[{\citenamefont{Starr et~al.}(2001)\citenamefont{Starr, Sastry, LaNave,
  Scala, Eugene~Stanley, and Sciortino}}]{Starr:01}
\bibinfo{author}{\bibfnamefont{F.~W.} \bibnamefont{Starr}},
  \bibinfo{author}{\bibfnamefont{S.}~\bibnamefont{Sastry}},
  \bibinfo{author}{\bibfnamefont{E.}~\bibnamefont{LaNave}},
  \bibinfo{author}{\bibfnamefont{A.}~\bibnamefont{Scala}},
  \bibinfo{author}{\bibfnamefont{H.}~\bibnamefont{Eugene~Stanley}},
  \bibnamefont{and}
  \bibinfo{author}{\bibfnamefont{F.}~\bibnamefont{Sciortino}},
  \bibinfo{journal}{Phys. Rev. E} \textbf{\bibinfo{volume}{63}},
  \bibinfo{pages}{041201} (\bibinfo{year}{2001}).

\bibitem[{\citenamefont{Derrida}(1981)}]{Derrida:81}
\bibinfo{author}{\bibfnamefont{B.}~\bibnamefont{Derrida}},
  \bibinfo{journal}{Phys. Rev. B} \textbf{\bibinfo{volume}{24}},
  \bibinfo{pages}{2613} (\bibinfo{year}{1981}).

\bibitem[{\citenamefont{Fischer and Hertz}(1999)}]{Fischer:99}
\bibinfo{author}{\bibfnamefont{K.~H.} \bibnamefont{Fischer}} \bibnamefont{and}
  \bibinfo{author}{\bibfnamefont{J.~A.} \bibnamefont{Hertz}},
  \emph{\bibinfo{title}{Spin Glasses}} (\bibinfo{publisher}{Cambridge
  University Press}, \bibinfo{year}{1999}).

\bibitem[{\citenamefont{Stillinger}(1988)}]{Stillinger:88}
\bibinfo{author}{\bibfnamefont{F.~H.} \bibnamefont{Stillinger}},
  \bibinfo{journal}{J. Chem. Phys.} \textbf{\bibinfo{volume}{88}},
  \bibinfo{pages}{7818} (\bibinfo{year}{1988}).

\bibitem[{\citenamefont{Shell and Debenedetti}(2004)}]{Shell:04}
\bibinfo{author}{\bibfnamefont{M.~S.} \bibnamefont{Shell}} \bibnamefont{and}
  \bibinfo{author}{\bibfnamefont{P.~G.} \bibnamefont{Debenedetti}},
  \bibinfo{journal}{Phys. Rev. E} \textbf{\bibinfo{volume}{69}},
  \bibinfo{pages}{051102} (\bibinfo{year}{2004}).

\bibitem[{\citenamefont{B\"uchner and Heuer}(1999)}]{Buchner:99}
\bibinfo{author}{\bibfnamefont{S.}~\bibnamefont{B\"uchner}} \bibnamefont{and}
  \bibinfo{author}{\bibfnamefont{A.}~\bibnamefont{Heuer}},
  \bibinfo{journal}{Phys. Rev. E} \textbf{\bibinfo{volume}{60}},
  \bibinfo{pages}{6507} (\bibinfo{year}{1999}).

\bibitem[{\citenamefont{Sciortino}(2005)}]{Sciortino:05}
\bibinfo{author}{\bibfnamefont{F.}~\bibnamefont{Sciortino}},
  \bibinfo{journal}{J. Stat. Mechanics} p. \bibinfo{pages}{05015}
  (\bibinfo{year}{2005}).

\bibitem[{\citenamefont{Moreno et~al.}(2006)\citenamefont{Moreno, Saika-Voivod,
  Zaccarelli, LaNave, Buldyrev, Tartaglia, and Sciortino}}]{Moreno:06}
\bibinfo{author}{\bibfnamefont{A.~J.} \bibnamefont{Moreno}},
  \bibinfo{author}{\bibfnamefont{I.}~\bibnamefont{Saika-Voivod}},
  \bibinfo{author}{\bibfnamefont{E.}~\bibnamefont{Zaccarelli}},
  \bibinfo{author}{\bibfnamefont{E.}~\bibnamefont{LaNave}},
  \bibinfo{author}{\bibfnamefont{S.~V.} \bibnamefont{Buldyrev}},
  \bibinfo{author}{\bibfnamefont{P.}~\bibnamefont{Tartaglia}},
  \bibnamefont{and}
  \bibinfo{author}{\bibfnamefont{F.}~\bibnamefont{Sciortino}},
  \bibinfo{journal}{J. Chem. Phys.} \textbf{\bibinfo{volume}{124}},
  \bibinfo{pages}{204509} (\bibinfo{year}{2006}).

\bibitem[{\citenamefont{Sastry}(2000)}]{SastryJP:00}
\bibinfo{author}{\bibfnamefont{S.}~\bibnamefont{Sastry}}, \bibinfo{journal}{J.
  Phys.: Condens. Matter} \textbf{\bibinfo{volume}{12}}, \bibinfo{pages}{6515}
  (\bibinfo{year}{2000}).

\bibitem[{\citenamefont{Matyushov}(2007)}]{DMpre:07}
\bibinfo{author}{\bibfnamefont{D.~V.} \bibnamefont{Matyushov}},
  \bibinfo{journal}{Phys. Rev. E} \textbf{\bibinfo{volume}{76}},
  \bibinfo{pages}{011511} (\bibinfo{year}{2007}).

\bibitem[{\citenamefont{Osipov et~al.}(1997)\citenamefont{Osipov, Teixeira, and
  da~Gama}}]{Osipov:97}
\bibinfo{author}{\bibfnamefont{M.~A.} \bibnamefont{Osipov}},
  \bibinfo{author}{\bibfnamefont{P.~I.~C.} \bibnamefont{Teixeira}},
  \bibnamefont{and} \bibinfo{author}{\bibfnamefont{M.~M.~T.}
  \bibnamefont{da~Gama}}, \bibinfo{journal}{J. Phys. A: Math. Gen.}
  \textbf{\bibinfo{volume}{30}}, \bibinfo{pages}{1953} (\bibinfo{year}{1997}).

\bibitem[{\citenamefont{Stillinger and Weber}(1985)}]{StillingerW:85}
\bibinfo{author}{\bibfnamefont{F.~H.} \bibnamefont{Stillinger}}
  \bibnamefont{and} \bibinfo{author}{\bibfnamefont{T.~A.} \bibnamefont{Weber}},
  \bibinfo{journal}{Phys. Rev. B} \textbf{\bibinfo{volume}{31}},
  \bibinfo{pages}{5262} (\bibinfo{year}{1985}).

\bibitem[{\citenamefont{Middleton and Wales}(2001)}]{Middleton:01}
\bibinfo{author}{\bibfnamefont{T.~F.} \bibnamefont{Middleton}}
  \bibnamefont{and} \bibinfo{author}{\bibfnamefont{D.~J.} \bibnamefont{Wales}},
  \bibinfo{journal}{Phys. Rev. B} \textbf{\bibinfo{volume}{64}},
  \bibinfo{pages}{024205} (\bibinfo{year}{2001}).

\bibitem[{\citenamefont{Allen and Tildesley}(1996)}]{Allen:96}
\bibinfo{author}{\bibfnamefont{M.~P.} \bibnamefont{Allen}} \bibnamefont{and}
  \bibinfo{author}{\bibfnamefont{D.~J.} \bibnamefont{Tildesley}},
  \emph{\bibinfo{title}{Computer Simulation of Liquids}}
  (\bibinfo{publisher}{Clarendon Press}, \bibinfo{address}{Oxford},
  \bibinfo{year}{1996}).

\bibitem[{\citenamefont{Brown and Clarke}(1984)}]{BrownClarke:84}
\bibinfo{author}{\bibfnamefont{D.}~\bibnamefont{Brown}} \bibnamefont{and}
  \bibinfo{author}{\bibfnamefont{J.~H.~R.} \bibnamefont{Clarke}},
  \bibinfo{journal}{Mol. Phys.} \textbf{\bibinfo{volume}{51}},
  \bibinfo{pages}{1243} (\bibinfo{year}{1984}).

\bibitem[{\citenamefont{Anderson et~al.}(1992)\citenamefont{Anderson, Bai,
  Bischof, Blackford, Demmel, Dongarra, Croz, Greenbaum, Hammarling, McKenney
  et~al.}}]{LAPACK}
\bibinfo{author}{\bibfnamefont{E.}~\bibnamefont{Anderson}},
  \bibinfo{author}{\bibfnamefont{Z.}~\bibnamefont{Bai}},
  \bibinfo{author}{\bibfnamefont{C.}~\bibnamefont{Bischof}},
  \bibinfo{author}{\bibfnamefont{S.}~\bibnamefont{Blackford}},
  \bibinfo{author}{\bibfnamefont{J.}~\bibnamefont{Demmel}},
  \bibinfo{author}{\bibfnamefont{J.}~\bibnamefont{Dongarra}},
  \bibinfo{author}{\bibfnamefont{J.~D.} \bibnamefont{Croz}},
  \bibinfo{author}{\bibfnamefont{A.}~\bibnamefont{Greenbaum}},
  \bibinfo{author}{\bibfnamefont{S.}~\bibnamefont{Hammarling}},
  \bibinfo{author}{\bibfnamefont{A.}~\bibnamefont{McKenney}},
  \bibnamefont{et~al.}, \emph{\bibinfo{title}{LAPACK users' guide}}
  (\bibinfo{publisher}{Philadelphia : Society for Industrial and Applied
  Mathematics}, \bibinfo{year}{1992}).

\bibitem[{\citenamefont{Angell}(2004)}]{Angell:04}
\bibinfo{author}{\bibfnamefont{C.~A.} \bibnamefont{Angell}},
  \bibinfo{journal}{J. Phys.: Condens. Matter} \textbf{\bibinfo{volume}{16}},
  \bibinfo{pages}{S5153} (\bibinfo{year}{2004}).

\bibitem[{\citenamefont{Mapes et~al.}(2006)\citenamefont{Mapes, Swallen, and
  Ediger}}]{Mapes:06}
\bibinfo{author}{\bibfnamefont{M.}~\bibnamefont{Mapes}},
  \bibinfo{author}{\bibfnamefont{S.}~\bibnamefont{Swallen}}, \bibnamefont{and}
  \bibinfo{author}{\bibfnamefont{M.}~\bibnamefont{Ediger}},
  \bibinfo{journal}{J.\ Phys.\ Chem.\ B} \textbf{\bibinfo{volume}{110}},
  \bibinfo{pages}{507} (\bibinfo{year}{2006}).

\bibitem[{\citenamefont{R\"ossler}(1990)}]{Rossler:90}
\bibinfo{author}{\bibfnamefont{E.}~\bibnamefont{R\"ossler}},
  \bibinfo{journal}{Phys. Rev. Lett.} \textbf{\bibinfo{volume}{65}},
  \bibinfo{pages}{1595} (\bibinfo{year}{1990}).

\bibitem[{\citenamefont{Giovambattista
  et~al.}(2004)\citenamefont{Giovambattista, Angell, Sciortino, and
  Stanley}}]{Giovambattista:04}
\bibinfo{author}{\bibfnamefont{N.}~\bibnamefont{Giovambattista}},
  \bibinfo{author}{\bibfnamefont{C.~A.} \bibnamefont{Angell}},
  \bibinfo{author}{\bibfnamefont{F.}~\bibnamefont{Sciortino}},
  \bibnamefont{and} \bibinfo{author}{\bibfnamefont{H.~E.}
  \bibnamefont{Stanley}}, \bibinfo{journal}{Phys. Rev. Lett.}
  \textbf{\bibinfo{volume}{93}}, \bibinfo{pages}{047801}
  (\bibinfo{year}{2004}).

\bibitem[{\citenamefont{Brebec et~al.}(1980)\citenamefont{Brebec, Seguin,
  Sella, Bevenot, and Martin}}]{Brebec:80}
\bibinfo{author}{\bibfnamefont{G.}~\bibnamefont{Brebec}},
  \bibinfo{author}{\bibfnamefont{R.}~\bibnamefont{Seguin}},
  \bibinfo{author}{\bibfnamefont{C.}~\bibnamefont{Sella}},
  \bibinfo{author}{\bibfnamefont{J.}~\bibnamefont{Bevenot}}, \bibnamefont{and}
  \bibinfo{author}{\bibfnamefont{J.~C.} \bibnamefont{Martin}},
  \bibinfo{journal}{Acta Metallurgica} \textbf{\bibinfo{volume}{28}},
  \bibinfo{pages}{327} (\bibinfo{year}{1980}).

\bibitem[{\citenamefont{Bartsch et~al.}(2006)\citenamefont{Bartsch, R{\"a}tzke,
  Faupel, and Meyer}}]{Bartsch:06}
\bibinfo{author}{\bibfnamefont{A.}~\bibnamefont{Bartsch}},
  \bibinfo{author}{\bibfnamefont{K.}~\bibnamefont{R{\"a}tzke}},
  \bibinfo{author}{\bibfnamefont{F.}~\bibnamefont{Faupel}}, \bibnamefont{and}
  \bibinfo{author}{\bibfnamefont{A.}~\bibnamefont{Meyer}},
  \bibinfo{journal}{Appl. Phys. Lett.} \textbf{\bibinfo{volume}{89}},
  \bibinfo{pages}{121917} (\bibinfo{year}{2006}).

\bibitem[{\citenamefont{Faupel et~al.}(2003)\citenamefont{Faupel, Frank, Macht,
  Mehrer, Naundorf, R\"atzke, Schober, Sharma, and Teichler}}]{Faupel:03}
\bibinfo{author}{\bibfnamefont{F.}~\bibnamefont{Faupel}},
  \bibinfo{author}{\bibfnamefont{W.}~\bibnamefont{Frank}},
  \bibinfo{author}{\bibfnamefont{M.-P.} \bibnamefont{Macht}},
  \bibinfo{author}{\bibfnamefont{H.}~\bibnamefont{Mehrer}},
  \bibinfo{author}{\bibfnamefont{V.}~\bibnamefont{Naundorf}},
  \bibinfo{author}{\bibfnamefont{K.}~\bibnamefont{R\"atzke}},
  \bibinfo{author}{\bibfnamefont{H.~R.} \bibnamefont{Schober}},
  \bibinfo{author}{\bibfnamefont{S.~K.} \bibnamefont{Sharma}},
  \bibnamefont{and} \bibinfo{author}{\bibfnamefont{H.}~\bibnamefont{Teichler}},
  \bibinfo{journal}{Rev. Mod. Phys.} \textbf{\bibinfo{volume}{75}},
  \bibinfo{pages}{237} (\bibinfo{year}{2003}).

\bibitem[{\citenamefont{Meyer et~al.}(1996)\citenamefont{Meyer, Wuttke, Petry,
  Peker, Bormann, Coddens, Kranich, Randl, and Schober}}]{Meyer:96}
\bibinfo{author}{\bibfnamefont{A.}~\bibnamefont{Meyer}},
  \bibinfo{author}{\bibfnamefont{J.}~\bibnamefont{Wuttke}},
  \bibinfo{author}{\bibfnamefont{W.}~\bibnamefont{Petry}},
  \bibinfo{author}{\bibfnamefont{A.}~\bibnamefont{Peker}},
  \bibinfo{author}{\bibfnamefont{R.}~\bibnamefont{Bormann}},
  \bibinfo{author}{\bibfnamefont{G.}~\bibnamefont{Coddens}},
  \bibinfo{author}{\bibfnamefont{L.}~\bibnamefont{Kranich}},
  \bibinfo{author}{\bibfnamefont{O.~G.} \bibnamefont{Randl}}, \bibnamefont{and}
  \bibinfo{author}{\bibfnamefont{H.}~\bibnamefont{Schober}},
  \bibinfo{journal}{Phys. Rev. B} \textbf{\bibinfo{volume}{53}},
  \bibinfo{pages}{12107} (\bibinfo{year}{1996}).

\bibitem[{\citenamefont{Chathoth et~al.}(2004)\citenamefont{Chathoth, Meyer,
  Koza, and Juranyi}}]{Chathoth:04}
\bibinfo{author}{\bibfnamefont{S.~M.} \bibnamefont{Chathoth}},
  \bibinfo{author}{\bibfnamefont{A.}~\bibnamefont{Meyer}},
  \bibinfo{author}{\bibfnamefont{M.~M.} \bibnamefont{Koza}}, \bibnamefont{and}
  \bibinfo{author}{\bibfnamefont{F.}~\bibnamefont{Juranyi}},
  \bibinfo{journal}{Appl.\ Phys.\ Lett.} \textbf{\bibinfo{volume}{85}},
  \bibinfo{pages}{4881} (\bibinfo{year}{2004}).

\bibitem[{\citenamefont{Sastry et~al.}(1998)\citenamefont{Sastry, Debenedetti,
  and Stllinger}}]{Sastry:98}
\bibinfo{author}{\bibfnamefont{S.}~\bibnamefont{Sastry}},
  \bibinfo{author}{\bibfnamefont{P.~G.} \bibnamefont{Debenedetti}},
  \bibnamefont{and} \bibinfo{author}{\bibfnamefont{F.~H.}
  \bibnamefont{Stllinger}}, \bibinfo{journal}{Nature}
  \textbf{\bibinfo{volume}{393}}, \bibinfo{pages}{554} (\bibinfo{year}{1998}).

\bibitem[{\citenamefont{Kuno et~al.}(2004)\citenamefont{Kuno, Shadowspeaker,
  Schroers, and Busch}}]{Kuno:04}
\bibinfo{author}{\bibfnamefont{M.}~\bibnamefont{Kuno}},
  \bibinfo{author}{\bibfnamefont{L.~A.} \bibnamefont{Shadowspeaker}},
  \bibinfo{author}{\bibfnamefont{J.}~\bibnamefont{Schroers}}, \bibnamefont{and}
  \bibinfo{author}{\bibfnamefont{R.}~\bibnamefont{Busch}},
  \bibinfo{journal}{Mat.Res. Soc. Proc.} \textbf{\bibinfo{volume}{806}},
  \bibinfo{pages}{227} (\bibinfo{year}{2004}).

\bibitem[{\citenamefont{Wang et~al.}(2006)\citenamefont{Wang, Angell, and
  Richert}}]{Richert:06}
\bibinfo{author}{\bibfnamefont{L.-M.} \bibnamefont{Wang}},
  \bibinfo{author}{\bibfnamefont{C.~A.} \bibnamefont{Angell}},
  \bibnamefont{and} \bibinfo{author}{\bibfnamefont{R.}~\bibnamefont{Richert}},
  \bibinfo{journal}{J. Chem. Phys.} \textbf{\bibinfo{volume}{125}},
  \bibinfo{pages}{074505} (\bibinfo{year}{2006}).

\end{thebibliography}

\end{document}